\documentclass[twocolumn,showpacs,amsmath,amssymb,prb]{revtex4} 
\usepackage{graphicx}

\def\rmd{{\rm d}}
\def\rme{{\rm e}}
\def\rmi{{i}}
\newcommand{\re}{\mathop{\rm Re}}
\newcommand{\im}{\mathop{\rm Im}}
\newcommand{\tr}{\mathop{\rm Tr}}
\newcommand{\hc}{\mathop{\rm h.c.}}
\newcommand{\ek}{\epsilon_{\bf k}}
\newcommand{\dk}{\Delta_{\bf k}}

\newcommand{\eref}[1]{(\ref{#1})}
\newcommand{\etal}{{\it et al.}}

\begin{document}

\title{On the Zero-Bias Conductance Peak in the Tunneling Spectroscopy}

\author{Shin-Tza Wu$^{1}$ and Chung-Yu Mou$^{1,2}$}
\affiliation{
1. Department of Physics, National Tsing Hua University, Hsinchu 30043,
Taiwan\\
2. National Center for Theoretical Sciences, P.O.Box 2-131, Hsinchu, Taiwan}

\date{\today}
\begin{abstract} 
A generalized method of image, incorporated with the non-equilibrium
Keldysh-Green's function formalism, is employed to investigate the
tunneling spectroscopy of hybrid systems in the configuration of
planar junction. In particular, tunneling spectroscopies of several
hybrid systems that exhibit zero-bias conductance peaks (ZBCP) are
examined. The well-known metal--$d$-wave superconductor (ND) junction
is first examined in detail. Both the evolution of the ZBCP versus
doping and the splitting of the ZBCP in magnetic fields are computed
in the framework of the slave-boson mean field theory.  Further
extension of our method to analyze other states shows that states with
particle-hole pairing, such as $d$-density wave and graphene sheet,
are all equivalent to a simple 1D model, which at the same time also
describes the polyacetylene. We provide the criteria for the emergence
of ZBCP. In particular, broken reflection symmetry at the microscopic
level is shown to be a {\em necessary condition} for ZBCP to occur.
\end{abstract}
\maketitle
\section{Introduction} 
Since the pioneering work of Giaever,\cite{giaever} the tunneling
measurement has been a major experimental method for investigations
into the electronic states of condensed matter systems.\cite{wolf} In
the simplest setup, a metal with {\it known} spectral property is made
in contact with a material X, forming an NX junction so that the
electronic states of X can be probed.  For many years, despite the
fact that many insights into the spectral properties of many states
have been gained from the differential conductance ($\rmd I/\rmd V$)
curves obtained from tunneling measurements, nonetheless, unlike many
other experiments, it is fair to say that there is no clear and solid
statement as to exactly what {\em bulk properties} are being probed in
tunneling measurements. For example, it is known that in neutron
scattering experiments, the neutron intensity is a measure of the
imaginary part of the bulk spin susceptibility, $\im \chi({\bf
k},\omega)$; no similar statement has ever been firmly established for
tunneling measurements.

The difficulty for establishing the relation between the tunneling
conductance and the bulk quantities can be traced back to the very
existence of the junction interface. It has been realized that the
presence of the interface can change the conductance curve
dramatically.  A well-known example is the zero-bias conductance peak
(ZBCP) observed in the tunneling spectra when X is a $d$-wave
superconductor (ND junction) in (110) direction.\cite{hu} The
appearance of the ZBCP is entirely tied up with the presence of the
interface and its orientations, and therefore can not be obtained by
simple calculations based on bulk density of states.
 
Recent theoretical analyses of the ZBCP have been mostly concentrated
on the ND junctions.  Furthermore, they are based largely on the
standard BTK theory.\cite{btk} In the continuum limit, analytic
expressions of the differential conductance for general orientations
of the interface were obtained. Numerical calculations were later
carried out for the BTK theory in the lattice
version.\cite{Tanuma,datta,cl} While these works have supplied
insights into the ZBCP, they are, however, specifically designed for
studying the ND junction.  Moreover, because the relation of the
conductance curve to the bulk quantities was not clearly manifested,
essentially the numerical computation had to be done individually for
each interface orientation. Another technical inconvenience is that
the BTK theory is a mean-field theory based on solving the mean-field
quasi-particle wavefunctions, it is thus difficult in this formulation
to take into account the effects of interaction systematically.  To
extend into the study of other systems, especially those with strong
correlations where almost all relevant models are on discrete
lattices, it is therefore an urgent need to have a formulation which
can go beyond the mean-field BTK formulations. As an illustration of
our approach, in this paper we will focus on mean-field analysis of
several tunneling problems. The effects of fluctuations and
interactions will be discussed elsewhere.

In this paper, we shall adopt an approach that is based on the
non-equilibrium Keldysh-Green's function formalism. In the lowest
order approximation, we will be able to express the differential
conductance entirely in terms of bulk Green's functions and include
the interface effects. Thus, the relation of the conductance curve to
the bulk quantities is clearly manifested.  The tunneling between N
and X will be treated as a perturbation, so that in the zeroth order
the Green's function is the mean-field {\em half-space} Green's
function that resides only on the semi-infinite plane and satisfies
the boundary conditions to be specified later.  Based on the
half-space Green's function, $g$, higher order corrections can be
systematically constructed.\cite{cl,luke,cuevas} In particular, a
class of infinite series in $g$, which consists of all elastic
tunneling processes in the perturbation theory, will be considered and
summed to all orders for calculating the current across the
junction.\cite{cl,caroli,yan} To fully take into account the
tight-binding nature of the problem, we shall employ discrete models
for both the materials N and X and the tunnel junctions. Thus the
essential quantity to be calculated is the half-space lattice Green's
function for the X state. In resemblance to the conventional method of
image, we express the half-space Green's function in terms of the bulk
Green's functions propagating from the real source and a fictitious
image source
\begin{eqnarray}
g = G_{\mbox{\rm real}} 
- G_{\mbox{\rm imag}} \times \alpha 
\label{green}
\end{eqnarray}
with the factor $\alpha$ accounting for the boundary conditions.  In
this picture the half-space Green's function is decomposed into two
parts: the real-source part comes solely from the bulk and hence
reveals purely the bulk properties, the image part contains all
interface effects which are encoded in the factor $\alpha$. In this
way, the interface effects are clearly identified in the course of the
analysis and one can pinpoint any departure from the bulk
property. 

The factor $\alpha$ can be expressed in terms of bulk Green's
function. Right on the interface, it is found 
\begin{eqnarray}
\alpha_0 = G^{-1}(d) G(-d). 
\label{alpha_0}
\end{eqnarray}
Here $d$ is an effective lattice constant whose precise meaning will
be explained in below. Clearly, the tunneling spectrum can be
classified according to whether the reflection symmetry is broken or
not. In the case when reflection symmetry is broken with respect to
the interface, one has $G(-d) \neq G(d)$, hence $\alpha_0$ is not
unity, possible zero modes may arise due to the presence of zeros in
the denominator of the left hand side. {\em The number of localized
zero mode is thus determined by the order of zeros in the bulk Green's
function $G(d)$.}  In the lowest order approximation, the differential
conductance is given by the local density of state at the interface
\begin{eqnarray}
\rmd I/\rmd V &\propto&
- \sum_{{\bf k}\sigma} \im\{g_0({\bf k},eV)\},
\end{eqnarray}
where $g_0$ is $g$ of Eq.~\eref{green} evaluated at the interface and
$e$ is the charge of an electron.  Since $\alpha_0$ can be expressed
entirely in terms of bulk Green's functions, this is then the relation
between the bulk quantities and the differential conductance alluded
to earlier.

This paper is organized as follows.  In Sec.~\ref{MOI}, we outline the
theoretical formulation and derive the generalized method of image for
discrete lattices. In Sec.~\ref{APP} this method is applied to the
study of tunneling spectroscopies for various systems. We first study
the ND junctions at various surface orientations and examine the
doping dependence of the ZBCP using mean-field slave boson theory. We
then study the effects of applied magnetic fields perpendicular to the
$ab$ plane. A one dimension model, based on the structure of
polyacetylene, is then studied in Sec.~\ref{POLY}. On the basis of
this model, we further apply this method to investigate tunneling into
$d$-density-wave states and graphite sheets. We conclude in
Sec.~\ref{fin} with some comments on the significance and further
applications of our formulation. The Appendix describes techniques for
deriving the current expressions for the tunnel junctions studied in
the text.

\section{Theoretical Formulation and Generalized method of image}
\label{MOI}
\subsection{Theoretical model}

\begin{figure}
\vspace*{-3mm}
\hspace*{10mm}
\includegraphics*[width=80mm]{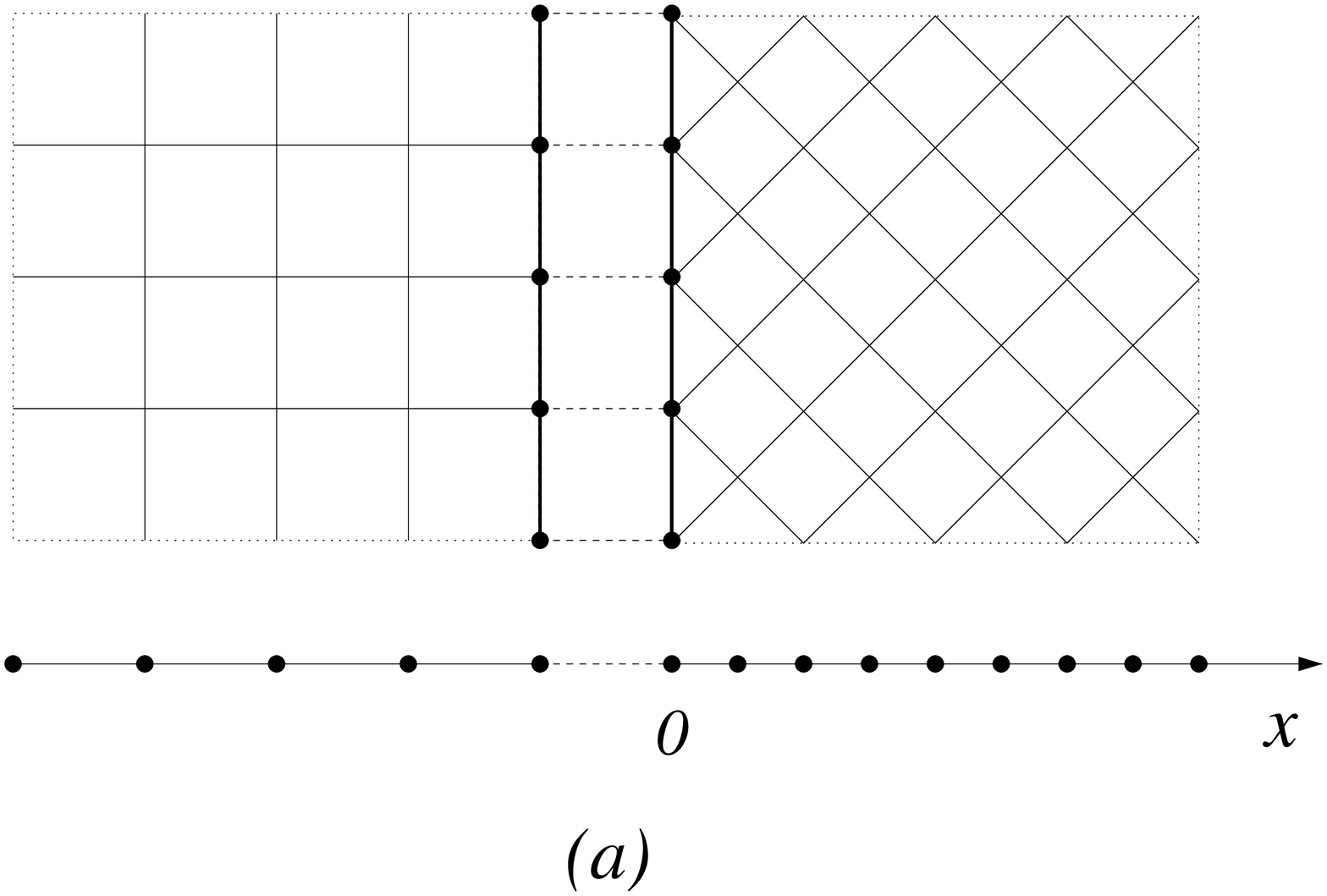}\\*[-20mm]
\hspace*{-20mm}
\includegraphics*[width=100mm]{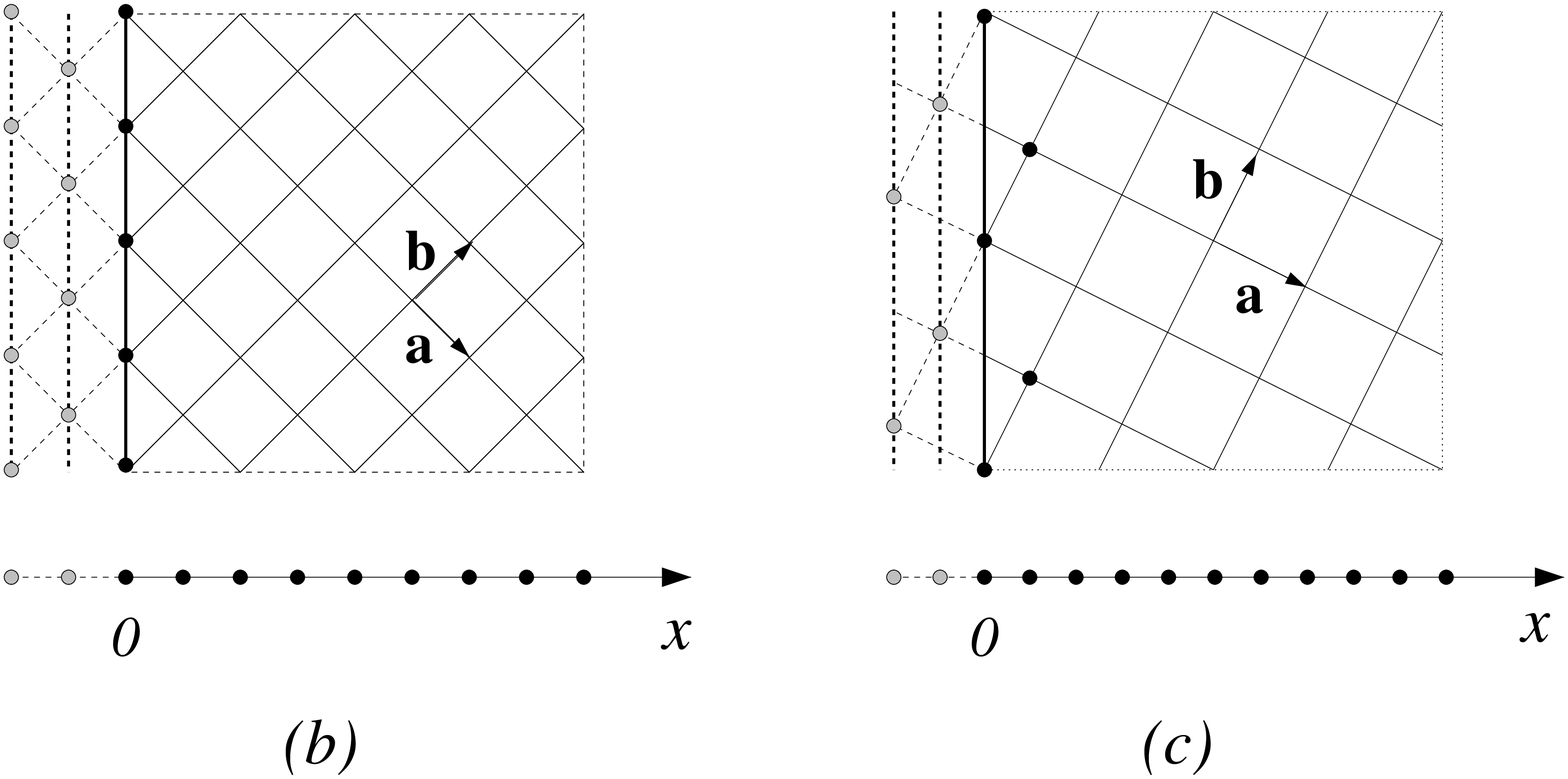}
\vspace*{-20mm}
\caption{\small $(a)$ A typical configuration for the tunnel 
junctions studied in this paper: a (100) lattice on the
left side connected to a (110) lattice on the right side. The dashed
lines between the two lattices indicate hopping due to the tunneling
Hamiltonian. The effective one dimensional lattices
obtained from Fourier transformation along the direction parallel to
the interface are shown in below. 
In $(b)$ and $(c)$ we show explicitly the hard walls of
semi-infinite lattices at (110) and (210) orientations. 
$a$ and $b$ indicate the crystalline axes. The
boundary at $x=0$ cuts off the lattice and hence there can be no
hopping across the boundary without the tunneling Hamiltonian. We
implement this boundary condition by setting up hard walls at all
lattice planes reachable by the boundary sites (filled circles in the
2D lattices) and requiring the Green's function to vanish over these
planes. In the (110) case, if only the nearest neighbor (n.n) hopping
is considered then only the first hard wall is needed; the second hard
wall imposes additional boundary conditions when there are next
nearest neighbor (n.n.n) hopping. For (210) orientation, however,
there are two hard walls even with only n.n. hopping. A third hard wall
would be needed if one considers n.n.n. hopping in the (210)
orientation.
\label{model} }
\end{figure}
We start by modeling the planar junction.
As illustrated in Fig.~\ref{model}, the tunnel junction consists of two
truncated two dimensional (2D) lattices connected through a tunnel
barrier, with the left half the normal (N) electrode ($-\infty <x\leq
-a_L$, $a_L$ is the lattice constant) and the right the test (X)
electrode ($0\leq x<\infty $).\cite{3D} We take the interface the $y$
direction. The total Hamiltonian of the system thus comprises two
parts: the Hamiltonian $H_0=H_L+H_R$ for the left and right
electrodes, and the tunneling Hamiltonian which connects the surface
points at $x=-a_L$ and $x=0$
\begin{eqnarray}
H_T= \sum_{y_l, y_r, \sigma} 
t(|y_l-y_r|) c_{l\sigma}^\dagger c_{r\sigma}^{} + \hc
\end{eqnarray}
Here $\sigma$ are spin indices, and $y_l$, $y_r$ are the
$y$-coordinates of the surface sites on the left and right electrodes;
$c_{l\sigma}$, $c_{r\sigma}$ are the corresponding electron
annihilation operators. $t$ is the tunneling amplitude whose magnitude
models the barrier height in the tunnel junction. Since all points
over the interface layers contribute to the tunneling process, one has
to sum over all interface sites. Suppose the chemical potentials on
the left and the right electrodes are $\mu_L$ and $\mu_R$,
respectively, the total grand Hamiltonian is
then given by
\begin{eqnarray}
K&=&(H_L-\mu_LN_L)+(H_R-\mu_RN_R)+H_T
\nonumber\\
&\equiv& K_0+H_T.  
\end{eqnarray}
The difference $(\mu_L-\mu_R)$ is fixed to be the voltage drop $eV$
across the junction.

To calculate the tunneling current, we choose the unperturbed state to
be the ground state of $K_0$ and adiabatically turn on $H_T$. In the
Heisenberg picture, the tunneling current is obtained from the time
rate of change of the particle number $N_L$ of the left
electrode\cite{mahan} (we set $\hbar = 1$ throughout)
\begin{eqnarray}
I(t) &=& + \rmi e \, \langle \; [ N_L, H_T ] \; \rangle 
\nonumber \\ 
&=&  + \rmi e 
\sum_{y_l,y_r,\sigma} \Big\{
t \langle c_{l\sigma}^\dagger c_{r\sigma}^{} \rangle 
-t^*  \langle c_{r\sigma}^\dagger c_{l\sigma}^{} \rangle  
\Big\} \, . 
\label{I_real}
\end{eqnarray}
The expectation values $\langle \cdots \rangle$ here represent the
ensemble average $\tr[Z^{-1} \exp{(-\beta K_0)} \cdots]$. 

In actual experiments, the normal metal on the left electrode could be
in any orientations, and the detail connection between the two
lattices may also cause complications in the tunneling spectroscopy. To
be definite, however, in our model we fix the lattice on the left side
at (100) orientation and connect its boundary sites to those of the
right at $x=0$ (Fig.~\ref{model}$(a)$). As one can observe easily, the
system is translational invariant along the interface direction with
period $a_L$, the lattice constant of the left side. We exploit this
symmetry by making a partial Fourier transformation along the
interface direction in Eq.~\eref{I_real} and arriving at 
\begin{eqnarray}
I(t) =  + \rmi e 
\sum_{k_y, \sigma}
t(k_y) \langle c_{l\sigma}^\dagger(k_y)
c_{r\sigma}^{}(k_y) \rangle 
+ {\rm h.c.} 
\label{I_ky}
\end{eqnarray}
The range of $k_y$ is determined by the periodicity of the interface
sites along $y$ direction, hence 
\begin{eqnarray}
-\pi/a_L < k_y \le \pi/a_L \,.
\label{ky_range}
\end{eqnarray} 
We emphasize that the problem is now effectively one dimensional: in
Eq.~\eref{I_ky} different $k_y$ modes are decoupled
completely. Moreover, only {\em surface} quantities are
involved. These are very appealing features especially for the
feasibility of our method of image, as we will discuss in the
following section.

In the Keldysh Green's function formulation the time evolution of the
density matrix can be formally solved as a closed time-ordered path
integral,\cite{keldysh} the expectation value $\langle
c_{l\sigma}^\dag(k_y) c_{r\sigma}^{}(k_y) \rangle$ in Eq.~\eref{I_ky}
is then related to the components of Keldysh's Green's functions over
the close time-path. One can then calculate perturbatively the average
current $I$ in terms of the zeroth order Green's function. Details of
this calculation can be found in Ref.~\onlinecite{cl} and an outline
is presented in the Appendix. Here an essential difference from
earlier work is that previously the Green's functions were obtained
through directly solving the equation of motion, while here we shall
make use of the method of image elucidated in the following
section. In this way the current approach is more general and
versatile, and can be easily applied to various hybrid systems.

\subsection{Generalized method of image}
In our scheme for the calculation of the tunneling current, the
building blocks are the zeroth order half-space Green's functions
(see the Appendix). Because in the zeroth order, lattices on the left and
right sides are disconnected, the Green's functions are defined only
for each semi-infinite plane. Therefore, lattice points on the
interface will have ``dangling bonds''. Effectively, as shown in
Figs.~\ref{model}$(b)$ and $(c)$, we are imposing hard-wall boundary
conditions at the end points of these dangling bonds. One thus
envisages a method of image similar to that in electrostatics.

In the usual practice, the method of image is done for the continuum
differential equations.  It is based on the principle of superposition
and the uniqueness of the solutions.\cite{jackson} When applying it to
the discrete lattice, one encounters the difficulty that the image
point to any source point $\bf{r}$, may not locate right at the
allowed lattice points.  To overcome this difficulty, we note that for
each semi-infinite lattice there is discrete translational invariance
along the surface direction, which we choose as the $y$
direction. Furthermore, since in the analysis of tunneling problems
each electrode is considered to be in steady states,
time-translational invariance is preserved in the individual
half-space. In the subsequent sections we shall fully exploit these
symmetries and thus will be concerned with the half-space Green's
function in its Fourier space representation $g(x,x';k_y;\omega)$,
which effectively propagates from $x'$ to $x$. For each $k_y$ and
$\omega$ one is therefore dealing with an effective one dimensional
(1D) system (Fig.~\ref{model}).

As a demonstration of the method, let us consider a 2D semi-infinite
square lattice with lattice constant $a$ extending over the region
$x\ge 0$ at orientation $(hk0)$. The hard-wall boundary condition
prescribes the half-space Green's functions to vanish over the hard
walls, which consist of all points where the boundary sites can reach
away from the bulk lattice (Figs.~\ref{model}$(b)$, $(c)$). For
general surface orientations $(hk0)$ and with only nearest-neighbor
(n.n.) hopping one can find that the number of hard walls is given by
$\max\{|h|,|k|\}$. Let us consider first the single hard-wall
configurations, which includes the $(100)$ and the $(110)$
orientations (when there is no next n.n hopping in the latter). As we
shall discuss later, the multi-hard-wall problem are simple
generalizations to the single hard-wall cases.

For single hard-wall case the only hard wall is located at $x=-d$,
where $d=a/\sqrt{h^2+k^2}$ is the spacing between two consecutive
$(hk0)$ planes. Since the Green's function must vanish on the hard
wall regardless the position of the source point $x'$, one imposes the
boundary condition
\begin{eqnarray}
g(-d,x';k_y;\omega) =0 \, .
\label{HWBC}
\end{eqnarray}
To implement the method of image, we construct the half-space Green's
function $g(x,x';k_y;\omega)$ from the full-space Green's function
$G(x,x';k_y;\omega)$ as
\begin{eqnarray}
g(x,x';k_y;\omega) = && G(x,x';k_y;\omega) 
\nonumber \\ 
&-& G(x,x_1';k_y;\omega) \alpha(x';k_y;\omega) ,
\label{g0}
\end{eqnarray}
where $x_1'=-2d-x'$ is the image point of the point source $x'$ with
respect to the hard wall $x=-d$. The Green's
function $G(x,x')$ describes direct propagation from the point source
to the point $x$, while $G(x,x_1')$ propagates from the image point to
$x$. The factor is determined by fitting the boundary condition \eref{HWBC}, we find
\begin{eqnarray}
\alpha(x')&=&G^{-1}(-d,-2d-x')G(-d,x')
\nonumber \\
&=&G^{-1}(d+x')G(-d-x'). 
\end{eqnarray}
Here and in the following we suppress the $k_y$ and $\omega$
dependence whenever no confusion would arise. In going from the
first to the second expressions above, we have used $G(x,x')=G(x-x')$,
namely that the full-space Green's functions are translational
invariant along the $x$ direction. However, this is not essential for
establishing the method of image. It is used here only for
brevity. For systems without translational symmetry along $x$
direction (such as the $d$-density-wave state to be discussed in
Sec.~\ref{DDW}), the following discussion still proceeds with only
minor modification.

From the half-space Green's function $g(x,x')$, one obtains the surface
Green's function $g_0$ by setting $x=x'=0$. In the Fourier space,
$g_0$ can be expressed by
\begin{eqnarray}
g_0 = \!\!\!\!\!\!\sum_{-\pi/d\le k_x<\pi/d} \!\!\!\!\!\!
G(k_x) \times [ 1 - \exp(2 \rmi k_x d) \alpha_0] \, .
\label{fourier}
\end{eqnarray}
Here $\alpha_0=\alpha(0)$ does not depends on $k_x$ and the sum over
$k_x$ extends over the first Brillouin zone of the effective 1D
lattice. The advantage of this formulation is clearly seen from
\eref{fourier}: the surface Green's function is obtained from
combinations of full-space Green's functions. For different surface
orientations, one simply rotates the full-space Green's function to
the appropriate angle. Furthermore, it is also clear that here we have
a scheme for studying the effects of interactions and fluctuations in
tunneling problems. Essentially one can take these effects into
account through the bulk Green's function. Here, however, we shall
concentrate on mean-field treatments and defer correlation effects to
a separate publication.

\begin{figure}
\includegraphics*[width=50mm]{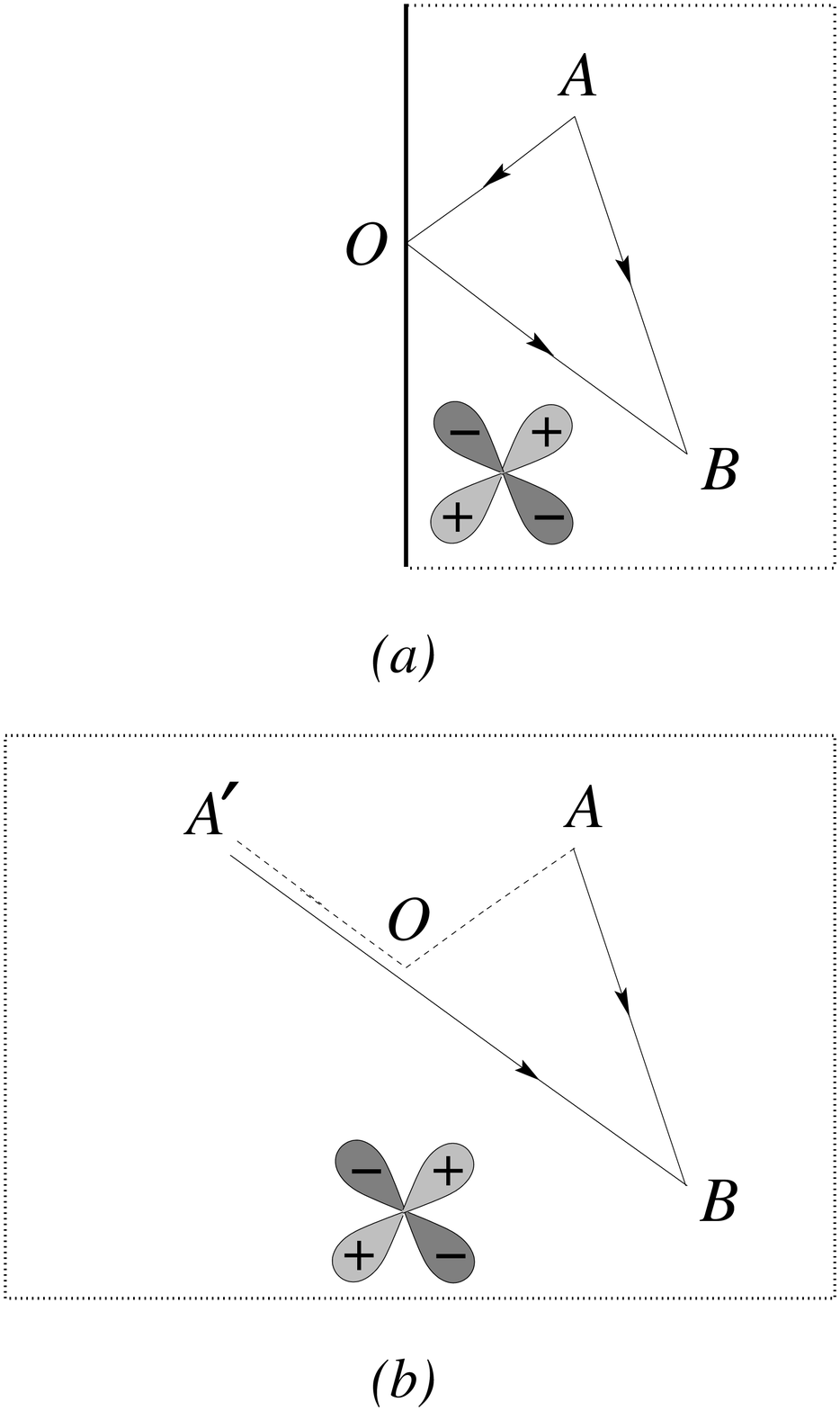}
\caption{\small The method of image applied to $d$-wave 
superconductors: the propagation $(a)$ from the source $A$ to the
point $B$ through the reflected path $AOB$ in the presence of a
hard-wall boundary can be replaced by $(b)$ a direct path $A'B$
emanating from a fictitious source at $A'$ where the boundary is
absent.
\label{image} }
\end{figure}

It is when dealing with lattices with an anisotropic order parameter
that one could most easily appreciate the power of the present
formulation. For instance in dealing with $d$-wave superconductors,
apart from fitting the boundary conditions \eref{HWBC}, $\alpha$ also
takes care of the different gap structures for propagation along the
reflected path and the fictitious path (such as $AO$ and $A'O$
depicted in Fig.~\ref{image}). In the presence of reflection symmetry
(such as an $s$-wave superconductor, or a $d$-wave superconductor at
$(100)$ orientation), since the gap structure as seen by these two
paths are identical, the full-space bulk Green's function possesses
the symmetry $G(d)=G(-d)$. Therefore $\alpha$ becomes (independent of
$k_y$ and $\omega$) universally equal to the identity matrix and
Eq.~\eref{fourier} reduces to the familiar form\cite{cl,yan}
\begin{eqnarray}
g_0 = \!\!\!\!\!\!\sum_{-\pi/d\le k_x<\pi/d}\!\!\!\!\!\!
G(k_x) \times 2 \sin^2(k_x d) \, .
\label{g_100}
\end{eqnarray}

For general orientations or when taking into account next nearest
neighbor hopping, as noted earlier, there could be more than one hard
walls. In these circumstances the surface Green's function must
satisfy the boundary condition that it vanishes on all these hard
walls simultaneously. This is a simple generalization of the single
hard-wall problem. For instance, let us consider the $(210)$ case with
n.n hopping: as depicted in Fig.~\ref{model}$(c)$ there are two hard
walls located at $x=-d$ and $-2d$, where $d=a/\sqrt{5}$. Analogous to
the single hard-wall problem, we write the half-space Green's function
\begin{eqnarray}
g(x,x') = G(x,x')\!-\! G(x,x_1') \alpha_1(x')
\!-\! G(x,x_2') \alpha_2(x')  
\label{g_210}
\end{eqnarray}
with $x_1'=-2d-x'$, $x_2'=-4d-x'$ being the location of the image
sources, and $\alpha_1$, $\alpha_2$ determined by the boundary
conditions
\begin{eqnarray}
g(-d,x') = 0 = g(-2d,x') \, .
\end{eqnarray}
In other words, for the point source at $x'$, each hard wall
``generates'' an image source on the other side of the surface and
introduces an $\alpha$ factor which accounts for the additional
boundary conditions. The half-space Green's function is a
superposition of contributions from the real and all image sources. To
obtain the surface Green's function, one again substitutes $x=x'=0$
into Eq.~\eref{g_210}.

Before proceeding to the applications in the following sections, we
comment that the present method is not restricted to square
lattices. In Sec.~\ref{GRAPH} we will apply this method to systems
involving honeycomb lattices. Indeed our generalized method of image
relies only on the possibility of reducing 2D lattices into 1D
structures through a Fourier transformation in the transverse
direction.

\section{Tunneling spectroscopy in hybrid systems}
\label{APP}
\subsection{Normal metal--$d$-wave Superconductors}
\label{NS}
We first study the $ab$-plane tunneling between a normal metal and a
$d$-wave superconductor. The superconductor occupies
the half-space $x>0$ and is described by the mean-field Hamiltonian
\begin{eqnarray}
H_R &=&-\!\!\! \sum_{\langle ij \rangle,\sigma} 
t_R c_{i\sigma }^\dag c_{j\sigma}^{}
-\!\!\! \sum_{\langle ij \rangle^{\prime },\sigma }
t_R' \,c_{i\sigma }^\dag c_{j\sigma}^{}
\nonumber\\
& & +\sum_{\langle ij \rangle}\Delta _{ij}
(c_{i\uparrow }c_{j\downarrow}-c_{i\downarrow}c_{j\uparrow })
+{\rm h.c.} \,,  \label{H_NS}
\end{eqnarray}
where $\langle ij \rangle$ denotes the nearest-neighbor (n.n.) bond,
$\langle ij \rangle'$ the next nearest neighbor (n.n.n.) bond, $t_R$
and $t_R'$ are hopping amplitudes between n.n and n.n.n sites,
respectively; $\Delta_{ij}$ is the mean-field pairing amplitude which
possesses the $d$-wave symmetry
\begin{eqnarray}
\Delta_{ij} = 
\left\{
       \begin{array}{cc}
       \Delta_0 &\mbox{\rm for ${\bf r}_j = {\bf r}_i + {\bf a}$,} \\
       -\Delta_0 &\mbox{\rm for ${\bf r}_j = {\bf r}_i + {\bf b}$.}
       \end{array}
\right. 
\label{Dij}
\end{eqnarray}
The normal metal on the left is modeled by a Hamiltonian similar to
$H_R$ but with only n.n hopping terms.

To obtain the corresponding 1D structure, we Fourier transform the
Hamiltonian along the $y$ direction. For example, at (110) orientation
if including only n.n. hopping $H_R$ becomes
\begin{eqnarray}
&&H_R =\sum_{x_i, k_y, \sigma } 
-2t_R\cos\!\left(\frac{k_y a}{\sqrt{2}}\right)
c_{i\sigma }^{\dagger}(k_y)c_{i+1\sigma}(k_y)  
\nonumber\\ && \hspace*{9mm}
+\sum_{x_i,k_y} 
2\rmi\Delta _0 \sin\!\left(\frac{k_y a}{\sqrt{2}}\right)
\left[ c_{i\uparrow }(k_y)c_{i+1\downarrow }(-k_y) 
\right. \nonumber\\ && 
\left. \hspace*{25mm}
+c_{i\downarrow}(-k_y)c_{i+1\uparrow }(k_y)\right] 
+{\rm h.c.} 
\label{H_NS_1D}
\end{eqnarray}
Here $c_i$ are the electron annihilation operators for the 1D lattice
at the $i$-th site (see Fig.~\ref{model}$(b)$) and $a$ is the lattice
constant of the original 2D lattice. The lattice constant of the 1D
lattice is identical to the distance $d$ between two consecutive
$(hk0)$ planes (cf. Fig.~\ref{model}). This 1D structure of the
problem is very helpful for us since each of the lattice planes
$(hk0)$ now becomes a point on the $x$ axis. This enables us to define
for each lattice site the corresponding image sites with respect to
the hard walls.\cite{wm}

It is convenient to use the Nambu notation which distinguishes
particle and hole components. We define the spinor
field-operator\cite{note_spinor}
\begin{eqnarray}
\Psi_i(k_y,t) = 
\left( \begin{array}{c}
        c_{i\uparrow}^{}(k_y,t)\\ 
        c_{i\downarrow}^\dag(-k_y,t)
        \end{array} 
\right) \, .
\label{NS_spinor}
\end{eqnarray}
The upper and the lower components of $\Psi_i$ correspond to the
particle and hole components, respectively. For (110) orientation one
can then write $H_R$ of \eref{H_NS_1D} in the form
\begin{eqnarray}
H_R &=& \sum_{x_i,k_y} \Big(
  \Psi_i^\dag H_{i,i+1} \Psi_{i+1} 
+ \Psi_i^\dag H_{i,i-1} \Psi_{i-1} \Big) 
\end{eqnarray}
with 
\begin{eqnarray}
H_{i,i\pm1} =
\left( \begin{array}{cc}
         -2t_R \cos(\frac{k_y a}{\sqrt{2}}) 
         & \pm 2\rmi\Delta_0 \sin(\frac{k_y a}{\sqrt{2}}) \\
         \pm 2\rmi\Delta_0 \sin(\frac{k_y a}{\sqrt{2}}) 
         & -2t_R \cos(\frac{k_y a}{\sqrt{2}})
       \end{array} 
\right) .
\label{H_NS_1D_nambu}
\end{eqnarray}

In order to apply the Keldysh formulation to calculating the tunneling
current, as detailed in the Appendix, the basic quantity one shall
need is the ({\rm bare}) half-space retarded Green's
function. According to our method of image this can be obtained from
superposition of the full-space Green's functions. Therefore our
remaining task is to find the {\em full-space} Green's function. For
general interface orientations $(hk0)$ all that we need is to rotate
the full-space Green's function to the appropriate angle and then
build up the half-space Green's function based on the recipe outlined
in Sec.~\ref{MOI}.

To find the full-space Green's function, we go over to the momentum
space and express the Hamiltonian \eref{H_NS} in Nambu's
representation
\begin{eqnarray}
H_R &=& \sum_{{\bf k},\sigma} 
\ek c_{{\bf k}\sigma}^\dag  c_{{\bf k}\sigma}^{}
+ \sum_{\bf k} (\dk c_{{\bf k} \uparrow}^\dag  
c_{-{\bf k} \downarrow}^\dag +
\dk^* c_{-{\bf k} \downarrow}  c_{{\bf k} \uparrow})
\nonumber \\
&=& \sum_{\bf k} 
\left( \begin{array}{cc}
              c_{{\bf k} \uparrow}^\dag  & c_{-{\bf k} \downarrow}^{}   
       \end{array} 
\right) \!
\left( \begin{array}{cc}
              \ek &  \dk \\
              \dk^* & -\ek 
       \end{array} 
\right) \!
\left( \begin{array}{c}
              c_{{\bf k} \uparrow}^{} \\ c_{-{\bf k} \downarrow}^\dag   
       \end{array} 
\right) \, .
\label{H_sc}
\end{eqnarray}
Here the quasiparticle dispersion $\ek$ and the gap function $\dk$ are
given by
\begin{eqnarray}
\ek &=& 
-2t_R [\cos({\bf k}\!\cdot\!{\bf a}) 
+ \cos({\bf k}\!\cdot\!{\bf b})] 
\nonumber \\&& 
-4t_R' \cos({\bf k}\!\cdot\!{\bf a}) \cos({\bf k}\!\cdot\!{\bf b}), 
\nonumber \\
\dk &=& -2 \Delta_0 [\cos({\bf k}\!\cdot\!{\bf a}) 
- \cos({\bf k}\!\cdot\!{\bf b})] \, .
\label{ek_dk}
\end{eqnarray} 
For $(hk0)$ orientation the lattice vectors ${\bf a}= a(\cos\theta,
-\sin\theta)$, ${\bf b}= a(\sin\theta, \cos\theta)$, where $a$ is the
lattice constant and $\theta$ is the angle between the $a$-axis and
the $x$-direction (thus $\tan\theta=k/h$). The Hamiltonian $H_R$ in
the form \eref{H_sc} is readily diagonalized, the quasiparticle
excitation energy is found to be $\pm E_{\bf k}=\pm
\sqrt{\ek^2+\dk^2}$.

The full-space retarded Green's function can be obtained from
\begin{eqnarray}
G(x_i,x_j) = \!\!\!\!\!\!\!\sum_{-\pi/d\le k_x<\pi/d}\!\!\!\!\!\!\!
G(k_x,k_y,\omega) \times \rme^{\rmi k_x (x_i-x_j)} \, , 
\label{G_bulk}
\end{eqnarray}
where $G(k_x,k_y,\omega) = [\omega + \rmi \eta -
\hat{H}_R(k_x,k_y,\omega)]^{-1}$, with $\hat{H}_R$ the matrix in the
second line of \eref{H_sc} and $\eta$ an infinitesimal positive
number. The half-space bare Green's function $g_0^r$ is then obtained
from the method of image. In the tunneling problem, the tunneling
Hamiltonian brings in tunneling events between the two sides of the
tunnel junctions which ``renormalize'' the half-space Green's
functions (see the Appendix). In the Keldysh formulation this is expressed
as a perturbation series which can be re-summed to all orders in the
tunneling amplitude $t$, yielding the renormalized half-space Green's
functions.\cite{caroli} With the assumption that the renormalized
advanced and retarded half-space Green's functions satisfy
\begin{eqnarray}
[g_{\alpha\beta}^r]^\dag = g_{\beta\alpha}^a  
\label{assume}
\end{eqnarray}
($\alpha,\beta=\{L,R\}$ labels the electrodes), we can express the
tunneling current as
\begin{eqnarray}
I = I_1 + I_2 + I_3 + I_A \, ,
\label{I_tot}
\end{eqnarray}
where
\begin{widetext}
\begin{eqnarray}
I_1 &=& \sum_{k_y} 4 \pi e \int_{-\infty}^\infty \rmd \omega 
\, t^2 [ f(\omega - eV) - f(\omega)] 
A_{L,11}(\omega - eV) A_{R,11}(\omega) 
|1+t g_{RL,11}^r(\omega)|^2 \, , \label{I1} 
\\ 
I_2 &=& \sum_{k_y} - 8 \pi e \int_{-\infty}^\infty \rmd \omega 
\, t^2 [ f(\omega - eV) - f(\omega)] 
A_{L,11}(\omega - eV) \re\{ A_{R,12}(\omega) 
[ t g_{LR,21}^a(\omega)(1+t g_{RL,11}^r(\omega))] \} 
\, , \label{I2}
\\ 
I_3 &=& \sum_{k_y} 4 \pi e \int_{-\infty}^\infty \rmd \omega 
\, t^4 [ f(\omega - eV) - f(\omega)] 
A_{L,11}(\omega - eV) A_{R,22}(\omega) 
|g_{RL,12}^r(\omega)|^2  \, , \label{I3}
\\ 
I_A &=& \sum_{k_y} 4 \pi e \int_{-\infty}^\infty \rmd \omega 
\, t^4 [ f(\omega - eV) - f(\omega + eV)] 
A_{L,11}(\omega - eV) A_{L,22}(\omega + eV) 
|g_{RR,12}^r(\omega)|^2 \, .  \label{Ia}
\end{eqnarray}
\end{widetext}
Here the range of the $k_y$ is given by \eref{ky_range},
\begin{eqnarray}
A_\alpha = \rmi/(2 \pi) 
(g_{0,\alpha \alpha}^r-{g_{0,\alpha \alpha}^r}^\dag)
\label{A}
\end{eqnarray}
are the spectral weight matrices for the electrode $\alpha=\{L,R\}$,
and $f(\omega)$ is the Fermi function (at zero temperature it is
simply the step function $\Theta(-\omega)$). The indices 1, 2 in the
Green's functions and the spectral weight matrices refer respectively
to the particle and the hole components in the Nambu
representation. $t=t(\omega, k_y)$ is the tunneling amplitude between
the two electrodes. It is remarkable that the expression for $I_2$
here generalizes that found in Ref.~\onlinecite{cl} and is applicable
to any interface orientation. For the special cases considered in
Ref.~\onlinecite{cl}, where the surface Green's functions are
symmetric (for $(100)$ orientation) or antisymmetric (for $(110)$
orientation), Eq.~\eref{I2} reproduces previous results. From these
formulas one can clearly identify the contributions from each channel
in the tunneling process. In particular, $I_1$ is the contribution
from single particle tunneling and $I_A$ the Andreev reflection (thus
$I_A$ depends on the particle and hole components of the spectral
weight matrix $A_L$).

\begin{figure}
\hspace*{-5mm}
\includegraphics*[width=90mm]{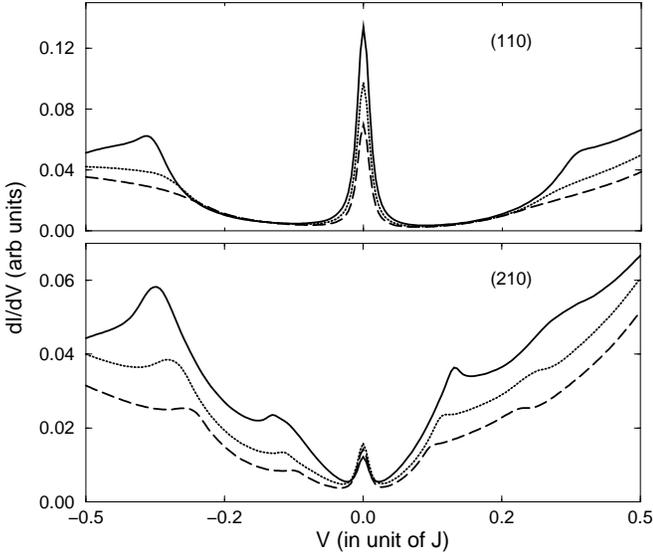} 
\caption{\small The total differential conductance for (110) and (210) 
interfaces at dopings $\delta=$ 0.08 (solid lines), 0.14 (dotted
lines), and 0.20 (dash lines). The weak link is modeled by the
interface hopping $t(\omega )= \exp (-\sqrt{(\omega _{0}-|\omega
|)/\Gamma })$. Here we use $\omega_{0}=11\Delta _{0}$ and $\Gamma
=\Delta _{0}$.
\label{dSC_dIdV} }
\end{figure}

We now present some of our results. Fig.~\ref{dSC_dIdV} shows the
tunneling spectra for (110) and (210) orientations at the doping
levels $\delta=0.08$, 0.14, and 0.20.  Here we study the doping
dependence by resorting to the mean-field slave boson theory for the
$t$-$t'$-$J$ model. The electron operators $c$ and $c^\dag$ are then
essentially the spinon operators and the Green's function for spinons
as well. The holons condense so that $\langle b \rangle
=\sqrt{\delta}$. The mean-field parameters $t_R$, $t_R'$, $\Delta_0$,
and the chemical potential $\mu_R$ for each doping are calculated
self-consistently.\cite{cl} It is obvious from Fig.~\ref{dSC_dIdV}
that the ZBCP is significantly reduced in the (210)
orientation. Interestingly, for (110) orientation the ZBCP decreases
upon increasing doping while for (210) case it grows and then falls
with doping. Another interesting feature in the tunneling spectra is
the subgap structures near $\pm 2\Delta_0$ in the (210) case. These
may have originated from resonances due to broken surface pairs,
resulting from the dangling bonds in (210) orientations.\cite{barash}
 
The ZBCP originates from zero-energy surface states (or the midgap
states) due to Andreev reflections. In our formulation these states
arise from singularities in the image contributions which manifest as
poles in the $\alpha$ factors. In the presence of a single hard wall,
the poles are determined by the zeros of the following factor when
$\eta=0$
\begin{eqnarray}
\beta(k_y)=\det[G(d;k_y,\omega=0)].
\label{beta}
\end{eqnarray}
This produces singular behavior in the Green's functions and results
in the ZBCP. In the (100) case, since $\alpha_0$ is simply the
identity matrix the surface Green's function \eref{g_100} is regular
at $\omega=0$ thus there is no ZBCP.

\subsection{Tunneling into current-carrying superconductors}
\label{CCS}
\begin{figure}
\includegraphics*[width=85mm]{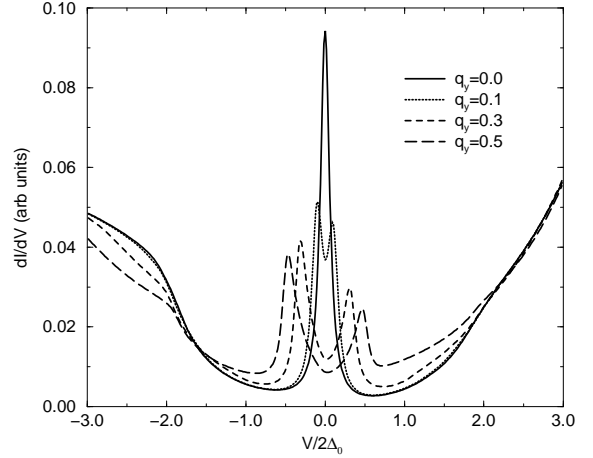} 
\caption{\small Splitting of the ZBCP for various values of $q_y$ 
(for $\delta = 0.16$).  
\label{ccs} }
\end{figure}

We now consider the $ab$-plane tunneling from normal metals to
current-carrying superconductors. In experiments one applies magnetic
field along the $c$-axis of the superconductor, so that a screening
current is generated over the $ab$ plane. When a quasiparticle tunnels
across the surface layer, it acquires additional energy from the
supercurrent. Thus the zero-energy surface state evolves in this case
into two surface states with non-zero energy. In the tunneling spectra
this appears as ``splitting'' of the ZBCP
(Fig.~\ref{ccs}). Fogelstr\"{o}m \etal~have analyzed splittings of the
zero-energy peak in the surface density of states under applied fields
in the continuum limit.\cite{fogels} Here we examine the tunneling
spectra base on our discrete model.

To marry with formulations in the previous section, we note that in
the presence of supercurrent the gap function is modified
as\cite{deGennes}
\begin{eqnarray}
\Delta_{ij} \rightarrow \Delta_{ij} 
\exp{(\rmi {\bf q} \cdot ({\bf r}_i + {\bf r}_j))} \, , 
\end{eqnarray}
where ${\bf q}=(0,q_y)$ is the superfluid momentum and is
proportional to the magnetic field. We shall assume
that tunneling events take place only within a shallow layer of order
about the penetration depth from the surface, so that $q_y$ is
approximately uniform in the region of our concern. This additional
phase can be absorbed into the electron operator by the transformation
$c_{i\sigma} \rightarrow c_{i\sigma}\exp(\rmi{\bf q}\cdot{\bf
r}_i)$. In Fourier space the Hamiltonian becomes
\begin{eqnarray}
H_R &=& \sum_{{\bf k},\sigma} 
\epsilon_{\bf k+q} c_{{\bf k},\sigma}^\dag  c_{{\bf k},\sigma}^{}
+ \sum_{\bf k} (\dk c_{{\bf k} \uparrow}^\dag  
c_{-{\bf k} \downarrow}^\dag +
\dk^* c_{-{\bf k} \downarrow}  c_{{\bf k} \uparrow})
\nonumber \\
&=& \sum_{\bf k} 
\left( \begin{array}{cc}
              c_{{\bf k} \uparrow}^\dag  & c_{-{\bf k} \downarrow}^{}   
       \end{array} 
\right) \!\!
\left( \begin{array}{cc}
              \epsilon_{\bf k+q} &  \dk \\
              \dk^* & -\epsilon_{\bf k-q} 
       \end{array} 
\right) \!\!
\left( \begin{array}{c}
              c_{{\bf k} \uparrow}^{} \\ c_{-{\bf k} \downarrow}^\dag   
       \end{array} 
\right) .
\label{H_ccs}
\end{eqnarray}
Here the quasiparticle dispersion $\ek$ and the gap function $\dk$ are
those of Eq.~\eref{ek_dk}. After diagonalizing $H_R$ above one finds
the quasiparticle excitation energy becomes
\begin{eqnarray}
E_{\bf k}^{(\pm)} 
= \left( \frac{\epsilon_{\bf k+q}-\epsilon_{\bf k-q}}{2} \right) 
\!\! \pm \!\sqrt{\left( \frac{\epsilon_{\bf k+q}
+\epsilon_{\bf k-q}}{2} \right)^2\!\! + \!\dk^2} \, .
\label{ekq}
\end{eqnarray}
The momentum space Green's function that is fed into
Eq.~\eref{fourier} is 
obtained in the same way: $G(k_x,k_y,\omega) = [\omega + \rmi
\eta-\hat{H}_R(k_x,k_y,\omega)]^{-1}$, with $\hat{H}_R$ the matrix in
the second line of \eref{H_ccs}.  

\begin{figure}
\includegraphics*[width=80mm]{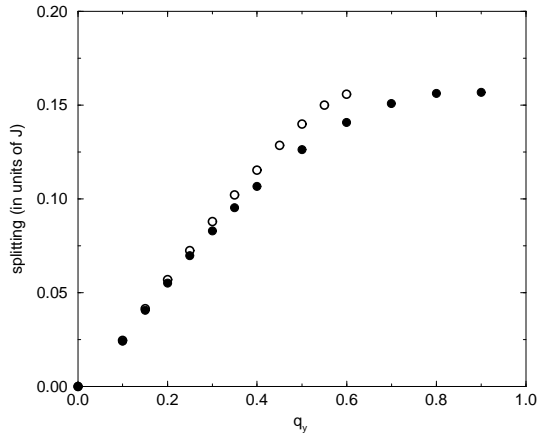} 
\caption{\small The dependence of splitting on magnetic field 
for doping $\delta = 0.12$. The empty and full symbols represent data 
calculated, respectively, with and without self-consistently taking into 
account the magnetic fields in solving the $t$-$t'$-$J$ slave boson 
mean-field equations. In the former case, the superconducting gap is 
strongly suppressed when $q_y\ge 0.65$, where difficulty in convergence 
of the mean-field solution arises.
\label{ccs_compare} }
\end{figure}

Fig.~\ref{ccs} shows typical tunneling spectra for the splitting of
the ZBCP when increasing $q_y$. Note that the slightly asymmetric
splitting originates from the particle-hole asymmetry in
$\epsilon_{\bf k}$. Fig.~\ref{ccs_compare} plots the magnitude of the
splitting versus the applied magnetic field for underdoped case. For
small ${\bf q}$ , the expansion of Eq~\eref{ekq} leads to linear
splitting in the lowest order terms: $ E_{\bf k}^{(\pm)}=\pm E_{\bf k}
+{\bf q}\cdot\frac{\partial\epsilon_{\bf k}}{\partial {\bf k}}$, where
$E_{\bf k} = \sqrt{ \epsilon_{\bf k}^2+\dk^2}$, as observed in small
applied fields.  For higher fields, one has to retain the full ${\bf
q}$ dependence, resulting in the bending of the splitting. This is
purely due to the lattice effect. Also shown in Fig.~\ref{ccs_compare}
are the results taking into account suppression of the superconducting
gap under magnetic fields self-consistently. The curve is seen to be
``pushed'' inwards while maintaining similar features. Note that
quantitative agreement with experimental
observations\cite{aprili,deutsch} can be obtained by fitting scales of
our results to the experimental data.  Nevertheless, we did not
observe any zero-field splitting at overdoping. This is in contrast
with the experiment of Ref.~\onlinecite{deutsch}. The mechanism
inducing this splitting might have eluded from our simple model.

\begin{figure}
\includegraphics*[width=90mm]{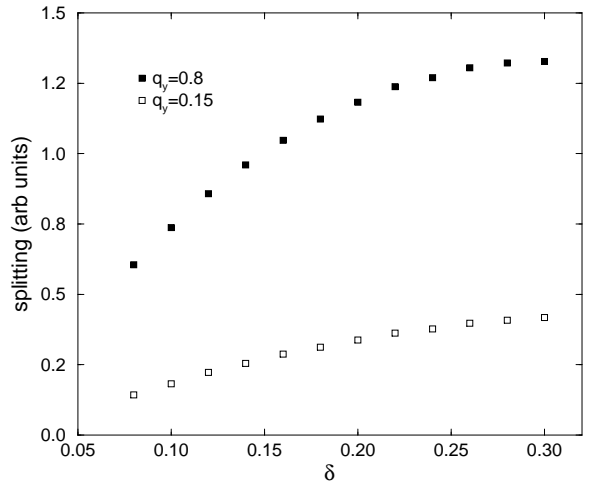} 
\caption{\small Splitting versus dopings for $q_y=0.15$ (open squares) 
and $0.8$ (solid squares). 
\label{ccs_doping} }
\end{figure}

The doping dependence of splitting is also shown in
Fig.~\ref{ccs_doping} for $q_y=0.15$ and $0.80$, which are
respectively in the linear and the saturated regimes in
Fig.~\ref{ccs_compare}. Note that the splitting increases with doping,
in agreement with Ref.~\onlinecite{deutsch}.

In passing we point out that the splitting depends sensitively on the
Fermi surface topology. Indeed for $\mu_R=0$ we find no splitting of
the ZBCP whatever the value of $q_y$ is. One can confirm this
analytically by making an asymptotic expansion of the Green's function
around $\omega = 0$. At $\mu_R=0$ one finds the conductance peak
invariantly stays at the zero bias.

\subsection{Polyacetylene} 
\label{POLY}
\begin{figure}
\includegraphics*[width=60mm]{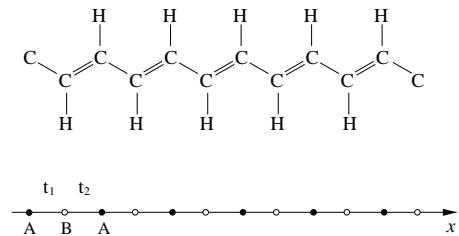}
\caption{\small The structure of polyacetylene and the 1D 
$t_1$-$t_2$ model. Filled and empty circles are lattice points over
the $A$ and $B$ sublattices.
\label{poly} }
\end{figure}
Up to this point, we have considered tunnel junctions with
superconducting test electrodes, where particles of opposite spins
form pairs. In this and the following sections we will consider
systems which exhibits particle-hole pairing over bipartite lattices.
To start with we shall consider first a simple 1D model based on the
structure of polyacetylene.\cite{su} This will turn out to be very helpful
for understanding results in the following sections. Most
importantly, it provides the criteria for the formation of midgap
states in semi-infinite bipartite systems.

The model we shall examine here is a 1D chain with alternating hopping
amplitudes $t_1$, $t_2$ as shown in Fig.~\ref{poly}. The separation
between the lattice points is taken to be a constant
$a$.\cite{note_poly} It is convenient to categorize the lattice points
into $A$ and $B$ sublattices and express the Hamiltonian for this
``$t_1$-$t_2$ model'' as
\begin{eqnarray}
H_R = \sum_{i_B,\sigma} 
-t_1 c_{i-1}^{A\dag} c_{i}^{B} 
- t_2  c_{i}^{B\dag} c_{i+1}^{A} + \hc
\label{H_poly}
\end{eqnarray}
Here $c_i^\alpha$ annihilates electrons over site $i$ on the
$\alpha=\{A, B\}$ sublattice (spin indices $\sigma$ will be omitted
throughout), and the sum run over sites $i$ in the $B$ sublattice
only.  Going over to momentum space, one finds
\begin{eqnarray}
H_R &=& \sum_{k,\sigma} 
\Lambda_k {c_k^A}^\dag c_k^B + \hc 
\nonumber \\
&=& \sum_{k, \sigma} 
\left( \begin{array}{cc}
              {c_k^A}^\dag  & {c_k^B}^\dag
       \end{array} 
\right) \!
\left( \begin{array}{cc}
              0 & \Lambda_k \\
              \Lambda_k^* & 0
       \end{array} 
\right) \!
\left( \begin{array}{c}
              c_k^A \\*[2mm]  c_k^B
       \end{array} 
\right) \, ,
\label{Hk_poly}
\end{eqnarray}
where 
\begin{eqnarray}
\Lambda_k &=& -t_1 \exp(\rmi ka) -t_2 \exp(-\rmi ka)
\nonumber \\
&=& -(t_1+t_2) \cos(ka) -\rmi (t_1-t_2) \sin(ka) \,. 
\end{eqnarray}
The Hamiltonian \eref{Hk_poly} can be diagonalized easily and the
quasiparticle excitation energy is found to be $\pm |\Lambda_k|$. Note
that the real part of $\Lambda_k$ is similar to the usual hopping
energy in one dimensional chain. Therefore, when
$\im\{\Lambda_k\}\propto (t_1-t_2)\neq 0$ a single particle excitation
gap opens at the chemical potential.

For semi-infinite chain, there are two possible configurations with
the terminating site being an $A$ or a $B$ sublattice point. In either
case we choose the boundary point the origin $x=0$ and construct the
surface Green's function utilizing the method of image
\begin{eqnarray}
g_0 = G(0,0) -G(0,-2a) G^{-1}(-a,-2a) G(-a,0) \, .
\label{g00_poly}
\end{eqnarray}
In this formula, the appropriate Green's functions should be used
depending the the type of the end point; for instance, in the case of
an $A$-type boundary even/odd sites are attributed to the $A$/$B$
sublattices. The retarded Green's function for infinite system is
\begin{eqnarray}
G(x_i^\alpha, t ;x_j^\beta, 0) = 
-\rmi \Theta(t) \left\langle 
\left\{ c_i^\alpha(t), {c_j^\beta}^\dag(0) \right\} 
\right\rangle 
\label{Gr_def}
\end{eqnarray}
for $x_i^\alpha$ over the $\alpha$ sublattices; the braces denote the
anti-commutators. After Fourier transformation in time, the Green's
function is obtained from
\begin{eqnarray}
G(x_i^\alpha,x_j^\beta;\omega) = 
\!\!\!\!\!\!\!\sum_{-\pi/2a\le k_x<\pi/2a}\!\!\!\!\!\!\!
G_{\alpha\beta}(k,\omega) \times 
\rme^{\rmi k(x_i^\alpha-x_j^\beta)} \, . 
\label{Gr_fourier}
\end{eqnarray}
Utilizing $\hat{H}_R$ the matrix in the second line of \eref{Hk_poly},
we write for brevity the momentum space Green's functions in matrix form
\begin{eqnarray}
G_{\alpha\beta}(k,\omega) &=&
[\omega + \rmi \eta - \hat{H}_R(k,\omega)]_{\alpha\beta}^{-1}
\nonumber \\
&=&\frac{1}{(\omega+\rmi \eta)^2- E_k^2}
\left(
	\begin{array}{cc} 
       	      \omega + \rmi \eta &  
              \Lambda_k     \\*[2mm]
              \Lambda_k^*   &
              \omega + \rmi \eta
        \end{array}
\right)_{\!\!\!\alpha\beta} \, .
\label{Gk_poly}
\end{eqnarray}
The matrix elements are assigned to the Green's functions according to
the convention in \eref{Hk_poly}, namely $\alpha=A,B$ corresponds to
$\alpha=1,2$ respectively.

Since in Eq.~\eref{g00_poly} the surface Green's function $g_0$ is
expressed as a combination of bulk Green's functions, the only
possible source of singular behavior in $g_0$ resides in the inverse
part $G^{-1}(-a,-2a)$. In other words, the existence of the
zero-energy mode depends on the behavior of $G(-a,-2a)$ at
$\omega=0$. This is analogous to the ND junctions where the ZBCP
results from the zeros of the determinant $\beta(k_y)$,
Eq.~\eref{beta}.

\begin{figure}
\includegraphics*[width=90mm]{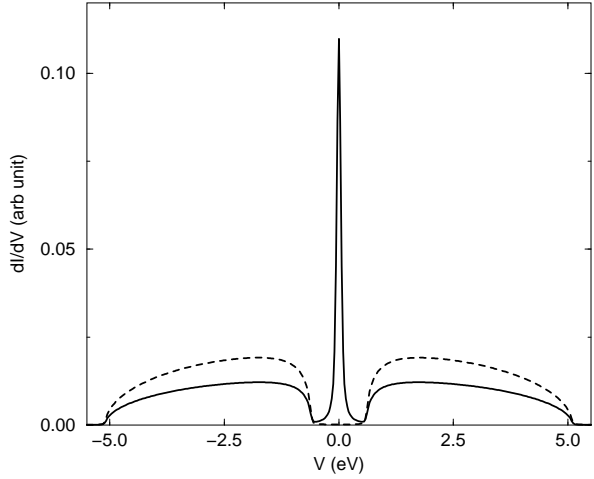}
\caption{\small Typical tunneling conductance curves for polyacetylene 
with $A$ type (solid line) and $B$ type (dashed line) end points. Here
we take $t_1=2.25\,$eV and $t_2=2.85\,$eV; thus the bandwidth is
$t_1+t_2=5.1\,$eV and the gap width $|t_1-t_2|=0.6\,$eV. The linear
chain on the left side has been taken a wideband material. In the
tunneling Hamiltonian $H_T$ we take $t=0.3$.
\label{poly_dIdV} }
\end{figure}

For example, in the case of $A$-type boundaries we find explicitly
(setting $\eta=0$)
\begin{eqnarray}
G_{BA}(-a,-2a;\omega\!=\!0)\!&=&\!\frac{a}{2\pi}
\int_{-\pi/2a}^{\pi/2a} \rmd k 
\frac{1}{t_1+t_2 \exp(-2\rmi ka)} 
\nonumber \\
&=& \left\{
          \begin{array}{cc}
           0     &\mbox{\rm if $t_1<t_2$,} \\
           1/2t_1 &\mbox{\rm if $t_2<t_1$.}
       \end{array}
\right. 
\label{criterion}
\end{eqnarray}
Here the index ``$BA$'' denotes the same meaning as in
\eref{Gk_poly} and is used for emphasizing the correct Green's function 
to be used. From \eref{criterion}, when $t_1<t_2$ a sharp singularity
in the surface Green's function $g_0$ arises at $\omega=0$ due to the
divergent factor $G_{BA}^{-1}$ in \eref{g00_poly}. On the other hand,
for $t_1>t_2$ since $G_{BA}^{-1}$ is finite at $\omega=0$, no singular
behavior in $g_0$ could occur there. Thus one expects ZBCP in the
former case while none in the latter. In the following sections we
shall see that this provides for 2D bipartite systems a criterion for
the range of transverse momenta where zero-energy states exist.  For
$B$-type boundary the analysis is identical, except an exchange in the
roles of $t_1$ and $t_2$. Therefore when the ZBCP shows up in an
$A$-type chain, it must be absent in a $B$-type chain, and vice
versa. This is shown in Fig.~\ref{poly_dIdV} for the case of
polyacetylene. The current expression here is identical to
Eq.~\eref{I_ddw} given in the following section, except the extra sum
over $k_y$ there.

\subsection{Normal metal--$d$-density wave states}
\label{DDW}
\begin{figure}
\includegraphics*[width=70mm]{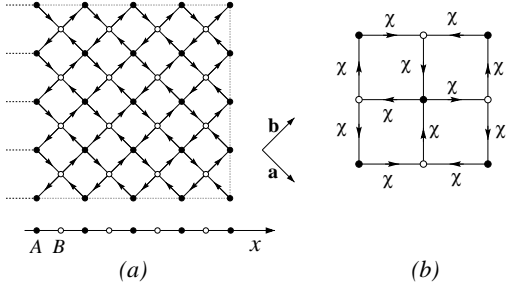} 
\caption{\small $(a)$ The configuration of a 2D square lattice with 
$d$-density-wave order at (110) orientation and the corresponding 1D
model. Filled and empty circles label the $A$ and $B$ sublattices; the
arrows indicate the directions of bond currents. Dashed lines extended
from the boundary sites depict coupling to the left electrode through
the tunneling Hamiltonian. $(b)$ shows explicitly the bond variables
in a doubled unit cell.
\label{ddw_lattice} }
\end{figure}

In underdoped cuprate superconductors, it is observed in experiments
that there are signatures of a ``partial'' gap well above the
superconducting temperature $T_c$. This anomalous regime in the phase
digram of the cuprate superconductors is thus termed the pseudogap
phase.\cite{timusk} Experiments also find that the pseudogap is
consistent with a $d$-wave structure. Recently Chakravarty
\etal~proposed that the pseudogap phase of the underdoped cuprate is
possibly the $d$-density-wave (DDW) state.\cite{chakravarty} It is
therefore of interest to examine the tunneling spectra of
normal-metal--$d$-density-wave (N-DDW) junctions.

The DDW state is characterized by the staggered flux in the elementary
plaquettes of the lattice. The bond currents circulating the unit cell
of the underlying square lattice break, among other symmetries, the
invariance of translation by one lattice spacing and lead to a
bipartite structure (Fig.~\ref{ddw_lattice}). Obviously, if the
interface cuts at (110) direction, the reflection symmetry is broken
-- in contrast to the (100) case. Therefore, we shall examine the
(110) direction with the following mean-field
Hamiltonian\cite{affleck}
\begin{eqnarray}
&&H_R = \sum_{i_B, \sigma} \left\{ 
\chi \left( c_{i+{\bf a}}^{A\dag} c_{i}^B 
+ c_{i-{\bf a}}^{A\dag} c_{i}^B \right) 
\right. \nonumber\\ && \hspace*{15mm} \left.
+ \chi^* \left( c_{i+{\bf b}}^{A^\dag} c_{i}^B 
+ c_{i-{\bf b}}^{A\dag} c_{i}^B \right) 
+ \hc  \right\} \, ,
\label{H_ddw}
\end{eqnarray}
where $c_i^\alpha$ annihilates an electron at site $i$ over the
$\alpha$ sublattice, and $\chi$ is the hopping amplitude on the bond
(Fig.~\ref{ddw_lattice}$(b)$).  Making Fourier transformation along
the $y$ direction in $H_R$, one finds
\begin{eqnarray}
H_R = \!\!\! \sum_{x_i^B, k_y, \sigma } \!\! \Big\{ 
\hspace*{-5mm}
&& \Lambda_{i,i-1}c_{i-1}^{A \dag}(k_y) c_{i}^B(k_y) 
\nonumber \\
&+& \Lambda_{i,i+1} c_{i}^{B \dag}(k_y) c_{i+1}^A(k_y) 
+ {\rm h.c.}  \Big\} \, ,
\label{H_ddw_1D}
\end{eqnarray}
where $\Lambda_{i,i\pm1}=2\re\{\chi\rme^{\pm\rmi k_ya/\sqrt{2}}\}$
with $a$ the lattice constant of the square lattice; $c_i^\alpha(k_y)$
is the electron annihilation operator for fixed $k_y$ at site
$x_i^\alpha$ over the 1D lattice. Going over to momentum space, one
finds, similar to \eref{Hk_poly}
\begin{eqnarray}
H_R = \sum_{{\bf k}, \sigma} 
\left( \begin{array}{cc}
              {c_{\bf k}^A}^\dag  & {c_{\bf k}^B}^\dag
       \end{array} 
\right) \!
\left( \begin{array}{cc}
              0 & \Lambda_{\bf k} \\
              \Lambda_{\bf k}^* & 0
       \end{array} 
\right) \!
\left( \begin{array}{c}
              c_{\bf k}^A \\*[2mm]  c_{\bf k}^B
       \end{array} 
\right) 
\label{Hk_ddw}
\end{eqnarray}
with $\Lambda_{\bf k} = \ek + \rmi \dk$. $\ek$ and $\dk$ are given by
Eq.~\eref{ek_dk} with $t_R=-\re\{\chi\}$, $\Delta_0=-\im\{\chi\}$, and
$t_R'=0$. The quasiparticle excitation energies is then obviously
$\pm E_{\bf k}=\pm|\Lambda_{\bf k}|=\pm\sqrt{\ek^2+\dk^2}$.

To find the tunneling current we apply again the Keldysh formulation
outlined in the Appendix. For fixed $k_y$ the full-space retarded
Green's function between the sites $x_i^\alpha$ and $x_j^\beta$
pertaining to the $\alpha$ and $\beta$ sublattices is given by
\begin{eqnarray}
\hspace*{-3mm}
G(x_i^\alpha, t ;x_j^\beta, 0;k_y) \!= \! 
- \rmi \Theta(t) \! \left\langle 
\left\{ c_i^\alpha(k_y,t), {c_j^\beta}^\dag(k_y,0) \right\} 
\right\rangle \! , 
\label{Gr_ddw}
\end{eqnarray}
which in Fourier space for fixed frequency $\omega$ becomes
\begin{eqnarray}
\hspace*{-3mm}
G(x_i^\alpha,x_j^\beta) =  
\!\!\!\!\!\!\!\sum_{-\pi/2d\le k_x<\pi/2d}\!\!\!\!\!\!\!
G_{\alpha\beta}(k_x,k_y,\omega) \times 
\rme^{\rmi k_x(x_i^\alpha-x_j^\beta)} .
\end{eqnarray}
Here $d=a/\sqrt{2}$ is the lattice spacing of the 1D lattice and we
have suppressed the $\omega$ and $k_y$ dependences on the left hand
side. $G_{\alpha\beta}$ has the same form given in Eq.~\eref{Gk_poly}
except that now $\alpha=A$ or $B$. 

\begin{figure}
\hspace*{-7mm}
\includegraphics*[width=90mm]{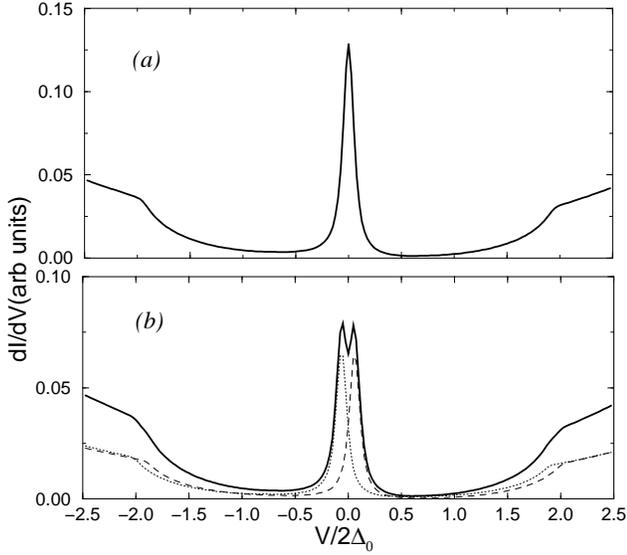} 
\caption{\small Typical conductance ($\rmd I/\rmd V$) curves for N-DDW 
junctions at (110) orientation in the $(a)$ absence and $(b)$ presence
of {\em in-plane} magnetic field. Here the boundary surface consists
of $A$ sublattice sites and
$\chi=(-t_R-\rmi\Delta_0)=(-0.447-0.1\rmi)$, $\eta=0.01$. Also shown
in $(b)$ are contributions from the spin-up (dashed line) and
spin-down (dotted line) components. The Zeeman splitting is here
$0.24\,\Delta_0$. The weak link is modeled by the same expression as
in Fig.~\ref{dSC_dIdV}.
\label{ddw_dIdV} }
\end{figure}
From $G_{\alpha\beta}$, the half-space surface Green's function 
is obtained again using the method of
image. The current expressions here, however, are distinct from those
of \eref{I1}--\eref{Ia}. Indeed since we are dealing with a single
component Green's function the calculation is much simpler than
previously. As shown in the Appendix, the current expression is here
\begin{eqnarray} 
\hspace*{-20mm}
&&I = \sum_{k_y,\sigma} 2 \pi e 
\int_{-\infty}^\infty \rmd\omega t^2 [f(\omega-eV) -f(\omega)] 
\nonumber \\ \hspace*{-20mm}
&& \hspace*{15mm}
\times A_L(\omega-eV) A_R(\omega)  |1+t g_{RL}^r(\omega)|^2 \, .
\label{I_ddw}
\end{eqnarray}
This is exactly the single-particle current $I_1$ of Eq.~\eref{I1} for
ND tunneling. There is no contribution from ``Andreev reflections'' in
N-DDW tunneling. This is due to the fact that in the DDW state the
pairing takes place between particles and holes of momenta ${\bf k}$
and ${\bf k+Q}$, with ${\bf Q}$ the nesting vector of 2D square
lattices. Thus the Andreev reflected particles are still electrons
whose response to the bias voltage are the same as the incident
particles; as a result their contributions to the tunneling current
cancel exactly. In the ND junction, however, a particle is Andreev
reflected as a hole, which behaves {\em oppositely} under applied
bias. Fig.~\ref{ddw_dIdV}$(a)$ shows a typical plot for differential
conductance versus voltage for N-DDW junctions. The conspicuous ZBCP
agrees with recent calculations done by Honerkamp and
Sigrist.\cite{honerkamp}

The reason for the ZBCP here can be understood on the basis of the
results in the previous section. Just like polyacetylene, the midgap
states arises when $g_0$ is singular due to the zeros in the Green's
function such as in Eq.~\eref{criterion}. For each $k_y$
Eq.~\eref{H_ddw_1D} resembles the $t_1$-$t_2$ model with
$t_1=-\Lambda_{i,i-1}$ and $t_2=-\Lambda_{i,i+1}$.  Therefore, for
example, for $A$-type boundary one expects midgap states for the range
of $k_y$ where
\begin{eqnarray}
\Lambda_{i,i-1}>\Lambda_{i,i+1} \quad {\rm or} \quad 
\im\{\chi\} \sin\left(\frac{k_ya}{\sqrt{2}}\right) > 0 \, .
\end{eqnarray}
Since here $\im\{\chi\}=-\Delta_0<0$, the above equation leads to
$-\sqrt{2}\pi/a<k_y<0$.

We have so far considered only the case of vanishing chemical
potential $\mu_R$ in the DDW state. At finite chemical potential the
grand Hamiltonian for the DDW is $K_R = H_R - \mu_R N_R$. Since the
number operator
\begin{eqnarray}
N_R = \sum_{{\bf k}\sigma} 
\left( \begin{array}{cc}
              {c_{\bf k}^A}^\dag  & {c_{\bf k}^B}^\dag
       \end{array} 
\right) \!
\left( \begin{array}{cc}
              1&0\\
              0&1
       \end{array} 
\right) \!
\left( \begin{array}{c}
              c_{\bf k}^A \\*[2mm] c_{\bf k}^B   
       \end{array} 
\right) \, .
\end{eqnarray}
Hence $-\mu_R N_R$ is diagonal and it simply shifts the excitation
energy $E_{\bf k}$ to $E_{\bf k}-\mu_R$. It is easy to check that this
change in the excitation energy induces a shift $\omega \rightarrow
\omega + \mu_R$ in the Green's function. This is in sharp
contrast with the ND case; there the chemical potential shifts the
quasiparticle energy $\ek \rightarrow \ek-\mu_R$ in the Green's
function but not the frequency. This results in the distinct behavior
of the ZBCP for ND and N-DDW junctions at finite $\mu_R$ .

For N-DDW junctions since $\omega \rightarrow \omega+\mu_R$ at finite
chemical potential, the conductance peak is shifted from zero bias to
the opposite value of the chemical potential $-\mu_R$. For ND
junctions, however, the midgap state stays at $\omega=0$ even at
finite chemical potential, thus the conductance peak always position
at zero bias (see Fig.~\ref{dSC_dIdV}). This shift has an obvious
implication: the peak will split due to the Zeeman splitting (see
Fig.~\ref{ddw_dIdV}). The orbital effects of magnetic fields can be
included by changing $\chi$ into $\chi \rme^{\rmi{\bf q}\cdot({\bf
r}_i - {\bf r}_j)}$ for any nearest neighbor sites $i$, $j$.  This
takes into account the current induced near the interface. Since under
this change both $\ek$ and $\dk$ undergo shifting of ${\bf k}$ by
${\bf q}$ which can be absorbed into the summation
of ${\bf k}$, 
the peak does not split. Therefore, the splitting of ZBCP
turn out the same for both in-plane and perpendicular magnetic fields.
This is in contrast to the ND junction where orbital effects dominate
for perpendicular fields.

\begin{figure}
\includegraphics*[width=90mm]{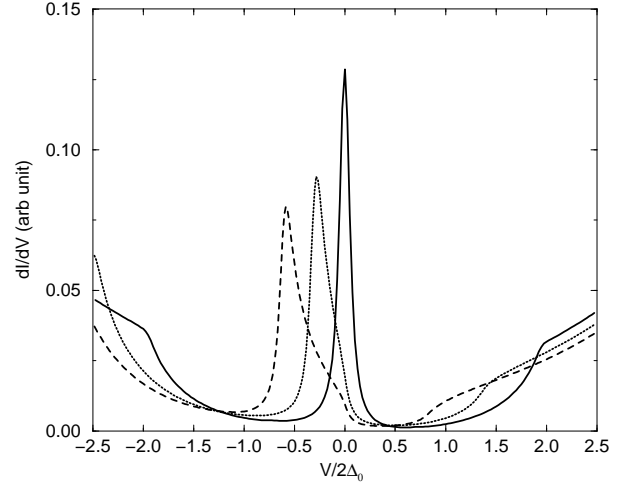}
\caption{\small Same as Fig.~\ref{ddw_dIdV}$(a)$ but with 
next n.n hopping amplitudes $t_R'=$ 0.0 (solid line), $-0.03$ (dotted
line), and $-0.06$ (dashed line).
\label{ddw_ttR} }
\end{figure}

In closing this section we note that since the next n.n term $-4t_R'
\cos({\bf k}\cdot{\bf a}) \cos({\bf k}\cdot{\bf b})$ couples only 
lattice sites within each sublattice, it introduces only diagonal
terms to \eref{Hk_ddw}. Therefore, similar to the chemical potential,
the next n.n terms cause the ZBCP and the spectrum to migrate when
$t_R'\neq 0$. This is displayed in Fig.~\ref{ddw_ttR}.

\subsection{Graphite sheets} 
\label{GRAPH}
\begin{figure}
\includegraphics*[width=90mm]{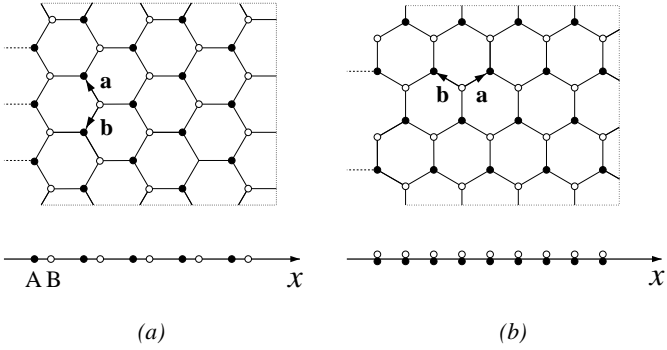}
\caption{\small Graphite sheets with $(a)$ zigzag and 
$(b)$ armchair boundaries and the corresponding 1D models after $k_y$
transformation. Filled and empty circles represent respectively the
$A$ and $B$ sublattices; ${\bf a}$, ${\bf b}$ are lattice vectors. The
dashed lines draw from the boundary sites indicate connections to the
left electrode through the tunneling Hamiltonian similar to
Fig.~\ref{model}$(a)$.
\label{graph} }
\end{figure}
So far we have considered systems involving only square lattices. As
commented in the end of Sec.~\ref{MOI}, our formulation is indeed
quite general and can be applied to any systems which can be projected
into 1D structures. As an example, we study in this section the
in-plane tunneling from a normal metal into semi-infinite graphite
sheets (NG junctions).

In the 2D graphite sheets, each carbon atom is bonded to its three n.n
via three electrons in the $sp^2$ orbital, forming a honeycomb
lattice. The fourth electron (the $\pi$ electron), which resides in
the $p_z$ orbital perpendicular to the 2D layer, is not bonded and is
free to hop from site to site. In the tight-binding limit the
Hamiltonian for the bulk graphite sheet is thus
\begin{eqnarray}
\hspace*{-3mm}
H_R = \!\!\sum_{i_B,\sigma}
-  \gamma_0  c_{i+{\bf   a}}^{A\dag} c_i^{B}  \!
-\!\gamma_1  c_{i+{\bf   b}}^{A\dag} c_{i}^{B}\!  
-\!\gamma_2  c_{i-{\bf a-b}}^{A\dag} c_{i}^{B}\!  
+ \hc 
\end{eqnarray}
Here the lattice is divided into $A$ and $B$ sublattices, and ${\bf
a}$, ${\bf b}$ are the lattice vectors illustrated in
Fig.~\ref{graph}.  $c_i^\alpha$ are electron annihilation operators at
site $i$ over the $\alpha$ sublattice, and $\gamma_i$ are the hopping
integrals. For simplicity we shall take $\gamma_0=\gamma_1=\gamma_2$
in the following. We will be interested in two orientations of the
lattice: one with zigzag and the other with armchair boundaries.

We first consider the zigzag case and choose the frame of coordinates
as shown in Fig.~\ref{graph}$(a)$. Fourier transformation in the
transverse direction leads to 1D Hamiltonian which resembles
\eref{H_poly}
\begin{eqnarray}
H_R = \!\!\!\!\! \sum_{x_i^B,k_y,\sigma} \!\!\!\!\!\!
-t_1 c_{i-1}^{A\dag}(k_y) c_i^{B}(k_y)\!
-\! t_2 c_i^{B\dag}(k_y) c_{i+1}^{A}(k_y)\! 
+\!\hc
\nonumber\\
\end{eqnarray}
with here 
\begin{eqnarray}
t_1 = 2\gamma_0 \cos(\frac{\sqrt{3}}{2}k_ya)  \quad {\rm and} \quad  
t_2=  \gamma_0 \, .
\label{t_eff}
\end{eqnarray}
Further $k_x$ transformation brings $H_R$ into the same form as
\eref{Hk_ddw} with 
\begin{eqnarray}
\Lambda_{\bf k} = - \gamma_0 
[\rme^{-\rmi k_x a} + 2\cos(\frac{\sqrt{3}}{2}k_ya) 
\rme^{\rmi k_xa/2} ] \, .
\end{eqnarray}

In applying the method of image, we note that the projected 1D lattice
for the zigzag case has alternating bond length, which breaks the
reflection symmetry and hence implies the possible existence of the
ZBCP. The alternating bond length, however, seems to cause difficulty
in locating the image point of an arbitrary source site. For instance,
the usual choice -- the mirror image -- does not always put the image
point right on the lattice. Nevertheless, since in 1D the hard wall
becomes a point, as long as the Green's function propagating from the
real source to the hard wall can be canceled by that from a
fictitious source so that the boundary condition is satisfied,
uniqueness of the half-space Green's function implies that the
location of the fictitious source can be chosen at will. Indeed, this
can be explicitly checked numerically.  To be definite, we shall place
the fictitious source at $x=-(3/2)a$ and apply the method of
image. The boundary condition $g(-a,x')=0$ for all $x'$ immediately
leads to
\begin{eqnarray}
\hspace*{-8mm}
g_0 &=& G_{AA}(0,0) 
\nonumber\\ \hspace*{-8mm} 
&-&G_{AA}(0,-3/2a) G_{BA}^{-1}(-a,-3/2a) G_{BA}(-a,0).
\end{eqnarray}
Here we have labelled the attributes of the lattice points explicitly
in the subscripts of the Green's functions. Just like polyacetylene,
the midgap states arises when $g_0$ is singular, namely at the zeros
of $G_{BA}(-a,-3/2a)$ when $\eta=0$. From \eref{t_eff} the
correspondence to the $t_1$-$t_2$ model indicates that midgap states
exist for $k_y$ which satisfy $\cos(k_y\sqrt{3}a/2)<1/2$, ie.~when
setting $\sqrt{3}a=1$
\begin{eqnarray}
-\pi \leq k_y < -2\pi/3 \quad {\rm and}
\quad 2\pi/3 < k_y \leq \pi \, .
\end{eqnarray}
This is exactly what is found in band structure 
calculations.\cite{fujita}
\begin{figure}
\includegraphics*[width=90mm]{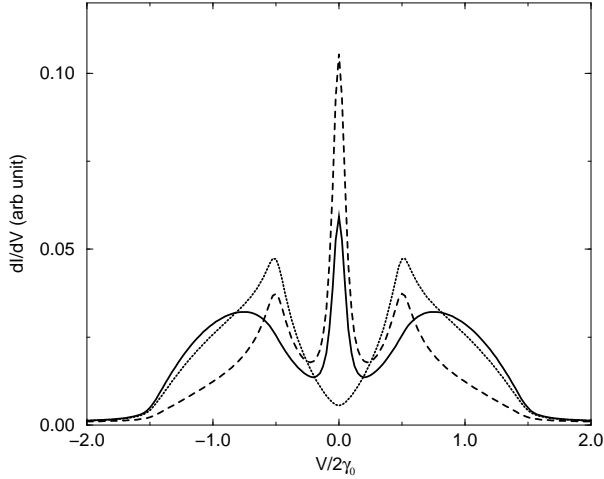}
\caption{\small Typical tunneling conductance curves for NG junctions 
with zigzag (solid line), bearded (dashed line), and armchair (dotted
line) boundaries. Here $\gamma_0=0.1$ and the left electrode has been
taken a wideband material. The weak link is modeled by the same
expression as in Fig.~\ref{dSC_dIdV} with here $\omega_0=11\gamma_0$
and $\Gamma=\gamma_0$.
\label{graph_dIdV} }
\end{figure}

For the zigzag orientation, apart from the zigzag boundary, there
could also be the ``bearded'' boundary where the surface layer
consists of $B$ sites. This is reminiscent of the case of $B$-type end
point of the $t_1$-$t_2$ model. Similar analysis as above can also
carry over here with minor change. We find in this case the zero
energy state arises from the range ($\sqrt{3}a=1$)
\begin{eqnarray}
-2\pi/3 < k_y < 2\pi/3 \, .
\end{eqnarray}
The current expression are the same as Eq.~\eref{I_ddw} for N-DDW
junctions. The corresponding tunneling spectra 
are solid and dashed lines shown in Fig.~\ref{graph_dIdV}.

Let us now consider the armchair case (Fig.~\ref{graph}$(b)$). 
After the Fourier
transformation along the interface, one finds
\begin{eqnarray}
H_R = \!\!\! \sum_{x_i^B,k_y,\sigma} \!\!\!
-\gamma_0 &\Big(& 
{c_i^B}^\dag(k_y) c_i^A(k_y) \rme^{-\rmi k_ya}
\nonumber \\*[-3mm]
&+& {c_i^B}^\dag(k_y) c_{i+1}^A(k_y) \rme^{\rmi k_ya/2}
\nonumber \\
&+& {c_i^B}^\dag(k_y) c_{i-1}^A(k_y) \rme^{\rmi k_ya/2}
+ \hc \Big)
\nonumber\\
\end{eqnarray}
Note that here for each site $x_i$ there are both $A$ and $B$ components as
shown in Fig.~\ref{graph}$(b)$. The propagation between $x_i$ and
$x_j$ thus compose of four components and the full-space Green's
function form a $2\times 2$ matrix
\begin{eqnarray}
G(x_i,x_j) = 
\left(
	\begin{array}{cc} 
       	      G(x_i^A,x_j^A) &  G(x_i^A,x_j^B) \\*[2mm]
              G(x_i^B,x_j^A) &  G(x_i^B,x_j^B)
        \end{array}
\right) \, . 
\end{eqnarray}
Further $k_x$ transformation yields the Hamiltonian \eref{Hk_ddw} with
here
\begin{eqnarray}
\Lambda_{\bf k} = - \gamma_0 
[\rme^{\rmi k_y a} + 2\cos(\frac{\sqrt{3}}{2}k_xa) 
\rme^{-\rmi k_y \frac{a}{2}}]\, .
\end{eqnarray}
Note that $\Lambda(-k_x,k_y)=\Lambda(k_x,k_y)$
implies that the reflection symmetry is not broken. 
Similar to the case of polyacetylene, 
in momentum space, $G({\bf k},\omega)$ has
exactly the same form as Eq.~\eref{Gk_poly}.
However, now the hard-wall
boundary condition becomes a matrix equation
\begin{eqnarray}
g(x_i^\alpha,x_j^\beta)\Big|_{x_i=-d}=0 
\qquad  \mbox{\rm for all $\alpha, \beta=\{A,B\}$} \, , 
\end{eqnarray}
where $d=(\sqrt{3}/2)a$ is the lattice constant of the projected 1D
lattice.  The surface Green's function then takes the form
\begin{eqnarray}
g_0 = G(0,0)-G(0,-2d) G^{-1}(-d,-2d) G(-d,0) \, .
\end{eqnarray}
Since translational symmetry is preserved in this 1D lattice, we have
$G(-d,-2d)=G(d)$ and $G(-d,0)=G(-d)$. Reflection symmetry implies
$G(d)=G(-d)$ and consequently
\begin{eqnarray}
g_0 = G(0)-G(2d)\, .
\label{g0_arm}
\end{eqnarray}
In this case, one thus expects no midgap states. 

Without loss of generality, we connect the $A$ sublattice to the left
side (Fig.~\ref{graph}$(b)$). The current expression is then the same
as Eq.~\eref{I_ddw}, where $g_0$ is replaced by the 11 (or $AA$)
component of the right hand side of Eq.~\eref{g0_arm}. The dotted line
of Fig.~\ref{graph_dIdV} shows the conductance curve for the armchair
case.

\section{Summary}
\label{fin}
In summary, the generalized method of image that we developed has
allowed us to deal with various tunneling problems in a unified
manner, with full tight-binding nature being taken into account.  In
particular, we have applied it in this paper to examine in-plane
tunneling spectra of normal metal--$d$-wave superconductor junctions,
with and without external magnetic fields, at arbitrary crystalline
orientations. The doping dependence of the ZBCPs is also studied
within the mean-field slave boson approach. Our results for the
splitting of ZBCP under applied magnetic field agrees well with recent
experiments. We further showed that tunneling into $d$-density-wave
state at (110) orientation should display a sharp conductance peak at
the chemical potential in the tunneling spectra. This peak will shift
away from the chemical potential if the next nearest neighbor hopping
$t_R'$ exists, which also offers a way to measure $t_R'$.  Under
in-plane magnetic fields, it also splits due to Zeeman splitting.
These provide signatures to be looked for in experiments, especially
in normal-metal--pseudogap-cuprate junctions for testing the proposal
of Ref.~\onlinecite{chakravarty}. As a demonstration of the general
applicability of our formulation, we further consider tunneling into
graphite sheets at the zigzag and armchair orientations. ZBCP is found
in the zigzag case while no ZBCP should be displayed in the armchair
case, consistent with findings in the study of graphite
ribbons.\cite{fujita} We analyze these results on the basis of the 1D
$t_1$-$t_2$ model and obtain the criteria for the emergence of the
ZBCP.

The merit of our formulation lies in two aspects. Firstly, it offers a
unified method for theoretical study of the tunneling
spectroscopy. Recently, applying this method, we discover a remarkable
even-odd effect in semi-infinite 1D chains: for hopping amplitudes
with even cycles there can be zero-energy localized edge-states, while
for those with odd cycles there is none.\cite{wm2} Secondly, as
already pointed out at the beginning of the paper, for the first time,
our method allows us to express what is being measured in tunneling
experiments in terms of bulk Green's function. For instance, in a
single hard-wall configuration, suppose the normal metal on the left
is a wideband material and the junction barrier is very high ($t\ll 1$
in Eqs.~\eref{I1}--\eref{Ia} and \eref{I_ddw}), tunneling experiments
essentially measure the quantity
\begin{eqnarray}
\rmd I/\rmd V &\propto& 
- \sum_{{\bf k}\sigma} \im\{g_0({\bf k},eV)\} 
\nonumber \\
&=& - \sum_{{\bf k}\sigma} \im\{
G({\bf k},eV) \times [ 1 - \exp(2 \rmi k_x d) \alpha_0] \} \, .
\nonumber \\ &&
\label{meaning}
\end{eqnarray}
Namely it probes the {\em surface} density of state, which is the bulk
Green's function modulated by the factor in the square bracket. In our
formulation the surface Green's function, and hence the surface
density of states, decomposes into two parts: one from the real source
and the other from the image source. The real-source part results
purely from the bulk and hence reveals faithfully the bulk property;
the image source part contains all surface effects and hence is
responsible for any complications. In the presence of the reflection
symmetry, we find that $\alpha_0=1$ and the image part contribute
another bulk term. Thus in this case the conductance curve exhibits
only the bulk property (with, however, the van Hove singularity
``rounded'' by a sine factor as in Eq.~\eref{g_100}). When the
reflection symmetry is broken, singular behavior may arise from the
image part. This singularity is contained in $\alpha_0$ and originates
from the zeros of the Green's function (Eq.~\eref{criterion}) or its
determinant when considering superconductors (Eq.~\eref{beta}). 

\begin{acknowledgments}
The authors would like to thank Profs. Sungkit Yip, Hsiu-Hau Lin,
T. K. Lee, Hu Xiao and C. C. Chi for useful discussions. This research was
supported by NSC of Taiwan.
\end{acknowledgments}

\appendix 

\renewcommand{\theequation}{\Alph{section}$\;\arabic{equation}$}
\section{Current expressions}
In this appendix we outline techniques for calculating the tunneling
currents for the ND and the N-DDW junctions. We start from the
expression
\begin{eqnarray}
I(t) =  + \rmi e 
\sum_{k_y, \sigma}
t(k_y) \langle c_{l\sigma}^\dagger(k_y)
c_{r\sigma}^{}(k_y) \rangle 
+ {\rm h.c.} 
\label{j}
\end{eqnarray}
In Keldysh's formulation of non-equilibrium Green's functions, the
expectation values $\langle c_{l\sigma}^\dag c_{r\sigma}^{}\rangle$ in
the above equation can be expressed as a perturbation series. Under
certain approximations one can resum this series to all orders in the
tunneling amplitude $t$. 

\subsection{ND junctions}
We consider first ND tunneling. In dealing with superconducting
phenomena it is convenient to use the Nambu representation which
explicitly distinguishes particles and holes by assigning them to
different components. One thus defines the spinor field-operators
\begin{eqnarray}
\Psi_\alpha(x_i,k_y,t) = 
\left( \begin{array}{c}
         \Psi_{\alpha,1} \\  \Psi_{\alpha,2}
        \end{array} 
\right)
= 
\left( \begin{array}{c}
        c_{\alpha\uparrow}^{}(x_i,k_y,t)\\ 
        c_{\alpha\downarrow}^\dag(x_i,-k_y,t)
        \end{array} 
\right) \, ,
\label{so}
\end{eqnarray}
where $\alpha=\{L,R\}$ labels the electrodes and the upper and lower
elements are associated with, respectively, electrons and holes. The
Keldysh non-equilibrium Green's functions are then defined
as\cite{yeyati,cuevas}
\begin{eqnarray}
\hspace*{-12mm}
&& g_{\alpha\beta,\mu\nu}^{-+}(x_i,t;x_j,t') 
= + \rmi \left\langle 
\Psi_{\beta,\nu}^\dag(x_j,t') 
\Psi_{\alpha,\mu}(x_i,t) \right\rangle \, ,
\\ \hspace*{-12mm}
&& g_{\alpha\beta,\mu\nu}^{+-}(x_i,t;x_j,t') 
= - \rmi \left\langle 
\Psi_{\alpha,\mu}(x_i,t) 
\Psi_{\beta,\nu}^\dag(x_j,t') \right\rangle \, .
\label{g_def}
\end{eqnarray}
For brevity we have suppressed the $k_y$ dependence. The Green's
functions here carry the left right indices $\alpha,\beta=\{L,R\}$,
the Nambu (spinor) indices $\mu,\nu =\{1,2\}$, and the Keldysh indices
$\{-,+\}$. For notational clarity we shall in the following frequently
omit irrelevant indices and keep track of only those related to our
discussion.

In this representation we define the tunneling matrix
\begin{eqnarray} 
\hat{t} \equiv t \, \tau_3 \, \sigma_3 \, 
\end{eqnarray}
where $\tau_3$ and $\sigma_3$ are the third Pauli matrices pertaining
to the Nambu space and the Keldysh space, respectively. In particular,
$\sigma_3$ is chosen so that in the Keldysh space
\begin{eqnarray}
\sigma_3^{--} = 1 = -\, \sigma_3^{++} \, ,
\quad {\rm and} \quad 
\sigma_3^{-+} = 0 = \,\,\, \sigma_3^{+-} \, ,
\nonumber
\end{eqnarray}
since we have assigned the forward time-path the ``$-$'' time axis,
and the return time-path the ``+'' time axis. In the following we will
consider only real valued $t$ and hence $t^*=t$.

The current expression \eref{j} can now be written as 
\begin{eqnarray}
\hspace*{-10mm}
&&I(t) = +e \sum_{k_y} \int_{-\infty}^\infty 
\frac{\rmd\omega}{2\pi} \,  t(k_y)  
\nonumber \\ \hspace*{-10mm}
&& \hspace*{5mm} \times \Big\{ 
\tr [g_{RL}^{-+}(x_{0},k_y,\omega)] 
- \tr[g_{LR}^{-+}(x_{0},k_y,\omega)] 
\Big\} \, .
\label{j_matrix}
\end{eqnarray}
where the trace is taken over the Nambu space and we have Fourier
transformed over the time variables. In the presence of particle-hole
symmetry the Nambu components 11 and 22 in the trace above contribute
equally. Therefore, the trace yields twice the contribution from the
11 component. Applying the Dyson equations\cite{keldysh}
\begin{eqnarray}
g = g_0 + g_0 \,\hat{t}\, g = g_0 + g \,\hat{t}\, g_0
\end{eqnarray}
and noting that $g_{0,RL}^{} = 0 = g_{0,LR}^{}$, as there is no
propagation between the two electrodes at the bare level, one finds
\begin{eqnarray}
g_{RL}^{-+} &=& \big[ g_0 + g \,\hat{t}\, g_0 \big]_{RL}^{-+}
\nonumber \\
&=& t(g_{RR}^{--} g_{0,LL}^{-+} - g_{RR}^{-+} g_{0,LL}^{++})
\label{gRL} 
\\
g_{LR}^{-+} &=& \big[g_0 + g_0 \,\hat{t}\, g\big]_{LR}^{-+}
\nonumber \\
&=& t(-g_{0,LL}^{-+} g_{RR}^{++} + g_{0,LL}^{--} g_{RR}^{-+})
\, . 
\label{gLR}
\end{eqnarray}
Since the components of Keldysh Green's functions has the property
$g^{++} + g^{--} = g^{-+} + g^{+-}$. Taking the difference between 
$g_{RL}^{-+}$ and $g_{LR}^{-+}$ using Eqs.~\eref{gRL} and \eref{gLR}
and inserting the result into \eref{j_matrix} we obtain
\begin{eqnarray} 
\hspace*{-15mm}
&& I = 2e \sum_{k_y} t^2
\int_{-\infty}^\infty \frac{\rmd \omega}{2\pi}
\Big\{ g_{0,LL,11}^{-+}(\omega - eV) g_{RR,11}^{+-}(\omega) 
\nonumber \\ \hspace*{-15mm}
&& \hspace*{30mm} 
- g_{0,LL,11}^{+-}(\omega - eV) g_{RR,11}^{-+}(\omega) \Big\} \, .
\label{j_exect}
\end{eqnarray}
Note that the frequency arguments of the bare Green's functions for
the left electrode $g_{0,LL}^{-+/+-}$ has been shifted due to the
applied bias $eV$ between the two sides ($\mu_L - \mu_R = eV$).  We
emphasize that at this stage Eq.~\eref{j_exect} is exact and our
remaining task is to express the exact Green's functions
$g_{RR,11}^{+-/-+}$ in terms of the uncoupled Green's functions $g_0$
via some approximations.

The Green's functions $g^{+-/-+}$ can be expressed in terms of the
bare Green's functions $g_0^{+-/-+}$ and the exact Green's functions
$g^{r,a}$ by means of the following equations\cite{keldysh}
\begin{eqnarray} \label{keldysh}
\hspace*{-5mm}
g^{+-/-+} (\omega) = 
[ 1 + g^r(\omega) \, \hat{t} \,] \, 
g_0^{+-/-+} [ \, \hat{t} \, g^a(\omega) + 1 ] \, .
\label{g_expand}
\end{eqnarray}
To fully reduce to bare quantities, one can express the exact Green's
functions $g^{r,a}$ in terms of the bare Green's functions $g_0^{r,a}$
by virtue of the Dyson equations
\begin{eqnarray} \label{dyson}
g^{r,a}(\omega) = g_0^{r,a}(\omega) 
+ g_0^{r,a}(\omega) \,\hat{t}\, g^{r,a}(\omega) \, .  
\end{eqnarray}
Ignoring many-body effects, we are able to resum the series and obtain
\begin{eqnarray}
\hspace*{-15mm}
&&g^{r,a}_{RR}(\omega) = 
{\cal T}_{RL}^{r,a}(\omega)
\: g^{r,a}_{0,RR}(\omega) \, , 
\label{grr} \\ \hspace*{-15mm}
&&g^{r,a}_{LR}(\omega) =
{\cal T}_{LR}^{r,a}(\omega)
\left[t\ g^{r,a}_{0,LL}(\omega-eV) \tau_3 g^{r,a}_{0,RR}(\omega)\right] \, ,
\\ \hspace*{-15mm}
&&g^{r,a}_{RL}(\omega) = 
{\cal T}_{RL}^{r,a}(\omega) 
\left[t\ g^{r,a}_{0,RR}(\omega) \tau_3 g^{r,a}_{0,LL}(\omega-eV)\right] \, ,
\\ \hspace*{-15mm}
&&g^{r,a}_{LL}(\omega) = 
{\cal T}_{LR}^{r,a}(\omega) 
\: g^{r,a}_{0,LL}(\omega-eV) \, ,
\label{grr_last}
\end{eqnarray}
where sum over tunneling processes of all orders is signified by the
factors
\begin{eqnarray*}
{\cal T}_{RL}^{r,a}(\omega) &=& 
\left[1 - t^2 g^{r,a}_{0,RR}(\omega) 
\tau_3 g^{r,a}_{0,LL}(\omega-eV) \tau_3 \right]^{-1} \, ,
\\
{\cal T}_{LR}^{r,a}(\omega) &=& 
\left[1 - t^2 g^{r,a}_{0,LL}(\omega-eV) 
\tau_3 g^{r,a}_{0,RR}(\omega) \tau_3 \right]^{-1} \, . 
\end{eqnarray*}
Since $\mu_L-\mu_R=eV$ all frequency arguments of $g_{0,LL}$ are to be
taken at $\omega-eV$. In arriving at the above equations we have used
for the normal electrode
\begin{eqnarray*}
g_{0,LL}(\omega-eV) =
\left( \begin{array}{cc}
        g_{0,LL,11}(\omega-eV) & 0\\
        0 & g_{0,LL,22}(\omega+eV)
       \end{array} 
\right) \! .
\end{eqnarray*}
Note that the 11 and the 22 elements of $g_{0,LL}$ represent particles
and holes, respectively, and hence their response to applied bias is
{\em opposite}. This is essential in giving rise to the Andreev
contributions in the tunneling current.

Incorporating Eqs.~\eref{grr}--\eref{grr_last} with \eref{keldysh},
one can thus obtain $g_{RR}^{-+/+-}$ and substitute back into
\eref{j_exect}. Finally we make use of the following relations
\begin{eqnarray}
g_0^{-+}(\omega) &=& 2 \pi \rmi f(\omega) \hat{A}(\omega) \, ,
\label{gnp} \\ 
g_0^{+-}(\omega) &=& - 2 \pi \rmi \, 
[1-f(\omega)] \, \hat{A}(\omega) \, ,
\label{gpn}
\end{eqnarray}
where $f(\omega)$ is the Fermi function and $\hat{A}(\omega)$ is the
spectral weight matrix given by \eref{A}. This leads us to 
the current expressions \eref{I1}--\eref{Ia}.

\subsection{N-DDW junctions}
We turn now to deriving the current expressions for N-DDW junctions
which is also applicable to NG junctions. We shall also show that in
this case Andreev-like processes do not contribute to the tunneling
current.  In the absence of external fields, spin degree of freedom
merely introduces a factor of two. Thus the spin indices $\sigma$ will
be omitted in the following.

We first define the Keldysh Green's functions similarly to
\eref{g_def} 
\begin{eqnarray}
\hspace*{-12mm}
&& g_{\alpha\beta}^{-+}(x_i,t;x_j,t') 
= + \rmi \left\langle 
c_{\beta}^\dag(x_j,t') c_{\alpha}(x_i,t) \right\rangle \, ,
\\ \hspace*{-12mm}
&& g_{\alpha\beta}^{+-}(x_i,t;x_j,t') 
= - \rmi \left\langle 
c_{\alpha}(x_i,t) c_{\beta}^\dag(x_j,t') \right\rangle \, .
\label{kds_ddw}
\end{eqnarray}
Here the subscripts $\alpha$, $\beta=\{R,L\}$ are labels for the
electrodes (not to be confused with the labels for sublattices in the
text). In terms of the Keldysh Green's functions the tunneling current
\eref{j} can be written
\begin{eqnarray}
I(t) = +e \sum_{l,r,\sigma} 
[ t \, g_{RL}^{-+}(r,l) - t^* \, g_{LR}^{-+}(l,r)] \, .
\end{eqnarray}
Similar to the previous section, the renormalized Green's functions
$g_{RL}^{}$ and $g_{LR}^{}$ can be expressed as combinations of the
bare Green's functions $g_{0,LL}^{}$ and the renormalized Green's
function $g_{RR}^{}$. This results in the exact formula
\begin{eqnarray} 
I = e \sum_{k_y,\sigma} t^2 
\int_{-\infty}^\infty \frac{\rmd \omega}{2\pi}
\Big\{ 
g_{0,LL}^{-+}(\omega - eV) g_{RR}^{+-}(\omega) 
\nonumber \\ 
-g_{0,LL}^{+-}(\omega - eV) g_{RR}^{-+}(\omega) \Big\} \, .
\label{I_ddw_total}
\end{eqnarray}
Note that here, unlike the previous section, the tunneling matrix is
\begin{eqnarray}
\hat{t} = t \sigma_3 \, ,
\end{eqnarray}
where $\sigma_3$ is the third Pauli matrix in the Keldysh space.

To proceed, as in the ND case, we apply \eref{g_expand} and expand the
exact Green's functions $g_{RR}^{+-/-+}$ in terms of the bare Green's
functions $g_0^{+-/-+}$ and the renormalized retarded and advanced
Green's functions $g^{r,a}$.  Again, ignoring many-body effects we can
express the renormalized Green's functions $g^{r,a}$ in terms of the
bare Green's functions $g_0^{r,a}$. Utilizing
\eref{gnp} and \eref{gpn}, one has
\begin{eqnarray}
\hspace*{-5mm}
&& g_{RR}^{-+} = 2 \pi \rmi 
[ f(\omega) M_R(\omega) + f(\omega-eV) M_L(\omega) ] 
\nonumber \\ \hspace*{-5mm}
&& g_{RR}^{+-} = -2 \pi \rmi 
[ (1-f(\omega)) M_R(\omega) + (1-f(\omega-eV)) M_L(\omega) ] 
\nonumber
\end{eqnarray}
with
\begin{eqnarray}
M_R(\omega) &=& A_R(\omega) |1+t g^r_{RL}(\omega)|^2 , \\
M_L(\omega) &=& t^2 A_L(\omega -eV) |g_{RR}^r(\omega)|^2 . 
\end{eqnarray}
In the last expressions we have used $g^r_{RL}=(g^a_{LR})^*$.  Note
that $M_L$ contains the spectral weight $A_L$ of the normal
electrode. It is associated with tunneling processes where a particle
is reflected back into the left side and at the same time a
particle-hole pair is transmitted into the right; this is reminiscent
of the Andreev channel in ND tunneling (cf.~the integrand in
\eref{Ia}). Substituting the above results into the current expression
\eref{I_ddw_total}, we obtain for the terms in the braces in the
integrand
\begin{eqnarray}
4 \pi^2 [f(\omega-eV) -f(\omega)] A_L(\omega-eV) M_R(\omega) \, ,
\end{eqnarray}
which leads to the current expression \eref{I_ddw}. 

It is remarkable that $M_L$ is canceled completely in the final
current expression, and hence no contribution from the Andreev process
remains in the tunneling current. The reason for this difference
between the ND and the N-DDW (and likewise NG) tunneling is that for
the former the bias voltage shifts the energy of particles and holes
in opposite directions, while for the latter particles from different
bands acquire the same shift under applied bias.


\begin{thebibliography}{99}
\bibitem{3D} The dimensionality of 2 is not essential; it is chosen 
here mainly for current experimental and theoretical interests. Our
formulation is equally applicable to 3D cases.
\bibitem{note_spinor} 
Here, for simplicity, we have omitted the labels for electrodes
$\{L,R\}$ (cf.~Eq.~\eref{so} in the Appendix).
\end{thebibliography}

\end{document}